\documentclass[journal]{IEEEtran} 
\usepackage{amsmath,amsfonts}
\usepackage{newtxtext,newtxmath}
\usepackage{algorithmic}
\usepackage{algorithm}
\usepackage{array}
\usepackage[caption=false,font=normalsize,labelfont=sf,textfont=sf]{subfig}
\usepackage{textcomp}
\usepackage{stfloats}
\usepackage{url}
\usepackage{verbatim}
\usepackage{graphicx}
\usepackage{cite}
\usepackage[
    colorlinks=true,
    linkcolor=blue,      
    citecolor=blue,      
    urlcolor=blue        
]{hyperref}
\usepackage{float}
\usepackage{mathtools}
\usepackage{environ}
\usepackage{pifont}
\usepackage{orcidlink}
\usepackage[dvipsnames]{xcolor}
\usepackage{calligra} 
\usepackage{mathrsfs} 
\usepackage{ragged2e}
\usepackage{multirow}
\usepackage{graphicx}
\usepackage{makecell}

\usepackage{orcidlink}  

\usepackage{acronym}

\acrodef{ook}[OOK]{on-off keying}
\acrodef{ask}[ASK]{amplitude shift keying}
\acrodef{fsk}[FSK]{frequency shift keying}
\acrodef{psk}[PSK]{phase shift keying}
\acrodef{bpsk}[BPSK]{binary PSK}
\acrodef{qpsk}[QPSK]{quadrature PSK}
\acrodef{dpsk}[DPSK]{differential PSK}
\acrodef{qam}[QAM]{quadrature amplitude modulation}
\acrodef{rqam}[RQAM]{rectangular QAM}
\acrodef{hqam}[HQAM]{hexagonal QAM}
\acrodef{xqam}[XQAM]{cross QAM}
\acrodef{ncfsk}[NCFSK]{non-coherent FSK}
\acrodef{lse}[LSE]{Laguerre series expansion}
\acrodef{clt}[CLT]{central limit theorem}

\acrodef{mrc}[MRC]{maximal ratio combining}
\acrodef{sc}[SC]{selection combining}
\acrodef{egc}[EGC]{equal gain combining}

\acrodef{pdf}[PDF]{probability density function}
\acrodef{cdf}[CDF]{cumulative distribution function}
\acrodef{mgf}[MGF]{moment generating function}
\acrodef{op}[OP]{outage probability}
\acrodef{cc}[CC]{channel capacity}
\acrodef{csi}[CSI]{channel state information}
\acrodef{se}[SE]{spectral efficiency}
\acrodef{aser}[ASER]{average symbol error rate}
\acrodef{aber}[ABER]{average bit error rate}
\acrodef{e2e}[e2e]{end-to-end}
\acrodef{snr}[SNR]{signal-to-noise ratio}
\acrodef{awgn}[AWGN]{additive white Gaussian noise}
\acrodef{acc}[ACC]{average channel capacity}
\acrodef{do}[DO]{diversity order}
\acrodef{cg}[CG]{coding gain}
\acrodef{cc}[CC]{channel capacity}
\acrodef{inid}[i.n.i.d.]{independent and non-identically distributed}
\acrodef{iid}[i.i.d.]{independent and identically distributed}

\acrodef{ffn}[FFN]{feedforward neural network}
\acrodef{dnn}[DNN]{Deep neural network}
\acrodef{mse}[MSE]{mean squared error}

\acrodef{bivfh}[BFHF]{bivariate Fox's H-function}
\acrodef{trivfh}[TFHF]{trivariate Fox's H-function}
\acrodef{multivfh}[MFHF]{multi-variate Fox's H-function}
\acrodef{ghz}[GHz]{gigahertz}

\acrodef{thz}[THz]{terahertz}
\acrodef{rf}[RF]{radio frequency}
\acrodef{fso}[FSO]{free space optics}
\acrodef{irs}[IRS]{intelligent reflecting surface}
\acrodef{mmwave}[mmWave]{millimeter wave}
\acrodef{b5g}[B5G]{beyond fifth-generation}
\acrodef{uav}[UAV]{unmanned aerial vehicle}
\acrodef{fspl}[FSPL]{free space path loss}
\acrodef{siso}[SISO]{single input single output}
\acrodef{uwb}[UWB]{ultra wideband}
\acrodef{noma}[NOMA]{non orthogonal multiple access}

\acrodef{e2e}[e2e]{end-to-end}
\acrodef{s}[S]{source node}
\acrodef{r}[R]{relay node}
\acrodef{d}[D]{destination node}
\acrodef{ap}[AP]{access point}
\acrodef{mus}[MUs]{mobile users}
\acrodef{mimo}[MIMO]{multiple input multiple output}
\acrodef{los}[LoS]{Line-of-Sight}
\acrodef{nlos}[NLoS]{non-LOS}
\acrodef{iot}[IoT]{Internet of Things}

\acrodef{awgn}[AWGN]{additive white Gaussian noise}

\newcommand{\refGamma}        {\cite[eq. (8.310.1)]{gradshteyn2014table}}  
\newcommand{\refMG}           {\cite[eq. (9.301)]{gradshteyn2014table}}    
\newcommand{\refFoxH}         {\cite[eq. (17)]{SRIVASTAVA1979191}}         
\newcommand{\refMVFoxH}       {\cite[eq. (A.1)]{Mathai2010}}               

\begin{document}

\title{Analytical and DNN-Aided Performance Evaluation of IRS-Assisted THz Communication Systems}

 \author{
    Soumendu Das\textsuperscript{\orcidlink{0000-0002-0167-095X}}, 
    Nagendra Kumar\textsuperscript{\orcidlink{0000-0002-7376-308X}}, \textit{Senior Member, IEEE},
    and Dharmendra Dixit\textsuperscript{\orcidlink{0000-0002-9845-7337}}, \textit{Senior Member, IEEE}%
    \thanks{S.~Das and N.~Kumar are with the Department of Electronics and Communication Engineering, National Institute of Technology Jamshedpur, Jamshedpur, India (e-mail: 2021rsec009@nitjsr.ac.in; kumar.nagendra86@gmail.com).}%
    \thanks{D.~Dixit (Corresponding Author) is with the Department of Electronics and Communication Engineering, Motilal Nehru National Institute of Technology Allahabad, Prayagraj, India (e-mail: dharmendradixit@mnnit.ac.in).}%
}

\maketitle

\begin{abstract}
This paper investigates the performance of an \ac{irs}-assisted \ac{thz} communication system, where the \ac{irs} facilitates connectivity between the source and destination nodes in the absence of a direct transmission path. The source–\ac{irs} and \ac{irs}–destination links are subject to various challenges, including atmospheric attenuation, asymmetric $\alpha$–$\mu$ distributed small-scale fading, and beam misalignment–induced pointing errors. The \ac{irs} link is characterized using the \ac{lse} approximation, while both the source–\ac{irs} and \ac{irs}–destination channels are modeled as \ac{iid} $\alpha$–$\mu$ fading channels. Furthermore, closed-form analytical expressions are derived for the \ac{op}, \ac{acc}, and \ac{aser} for \ac{rqam} and \ac{hqam} schemes over the \ac{e2e} link. The impact of random co-phasing and phase quantization errors are also examined. In addition to the theoretical analysis, deep neural network-based frameworks are developed to predict key performance metrics, facilitating fast and accurate system evaluation without computationally intensive analytical computations. Moreover, the asymptotic analysis in the high-\ac{snr} regime yields closed-form expressions for coding gain and diversity order, providing further insights into performance trends. Finally, Monte Carlo simulations also validate the theoretical formulations and present a comprehensive assessment of system behavior under practical conditions.
\end{abstract}


\acresetall
\section{Introduction}
The demand for ultra-high-speed, low-latency wireless networks has led to the exploration of the \ac{thz} band (0.1–10 \ac{thz}) as a promising solution for 6G and beyond. \ac{thz} communication offers ultra-wide bandwidth, provides Tbps data rates while accommodating a massive number of users \cite{akyildiz2022terahertz}. However, severe atmospheric attenuation due to molecular absorption, particularly by water vapor, limits its propagation distance. To mitigate path loss, highly directional \ac{thz} beams are employed, but this introduces challenges such as pointing errors arising from antenna misalignment \cite{boulogeorgos2020outage}. Additionally, shadowing caused by obstacles necessitates an effective means of maintaining communication. \Ac{irs} have emerged as a viable solution to mitigate blockages and enhance coverage in \ac{thz} networks. By dynamically adjusting the phase shifts of reflected signals, \ac{irs} can extend communication range without requiring additional power.

\subsection{Related Literature}
Recent research is therefore focusing on \ac{irs}-assisted \ac{thz} band communication. In \cite{M1}, the authors proposed an \ac{uav}-aided \ac{thz} band communication system, optimizing user grouping and \ac{irs} phase shifts using a deep reinforcement learning.
In \cite{M2}, a Beyond Diagonal-\ac{irs}-assisted \ac{thz} system was proposed, optimizing hybrid beamforming and phase shifts to enhance sum rate.
In \cite{M3}, a fast beam training and alignment scheme was proposed for \ac{irs}-assisted \ac{thz}/\ac{mmwave} systems, optimizing beamforming gain and beam training success rates in both LOS and NLOS scenarios.
In \cite{M4}, a spherical wave channel model for \ac{irs}-assisted \ac{thz} systems was proposed, analyzing power gain, energy efficiency, and beamforming performance, demonstrating the superiority of near-field beamfocusing over conventional far-field beamforming.
The authors in \cite{M5} presented a reliability and security analysis of wireless systems over cascaded $\alpha-\mu$ fading channels.
The authors in \cite{M6} presented a performance analysis of an \ac{irs}-assisted \ac{thz} communication system, deriving exact \ac{op}, \ac{aber}, and \ac{acc} expressions using multivariate Fox’s H-function.
The authors in \cite{M7} analyzed \ac{irs}-assisted \ac{uwb} \ac{thz} communications, deriving upper bounds on achievable rate and evaluating phase-shift optimization strategies to mitigate beam split effects and enhance system performance.
The authors in \cite{M8} proposed a time delay-based \ac{irs}-aided \ac{thz} communication scheme to mitigate beam squint, formulating a joint beamforming optimization problem and evaluating weighted sum rate performance using alternating optimization and convex optimization techniques.
The authors in \cite{M9} proposed a space-orthogonal precoding scheme for multi-\ac{irs}-aided multi-user \ac{thz} \ac{mimo} systems, evaluating weighted sum-rate performance under various beamforming techniques.
The authors in \cite{M10} analyzed the performance of a low-complexity algorithm for user association in multi-\ac{irs}-aided \ac{thz} networks.
The authors in \cite{M11} analyzed the performance of an \ac{irs}-assisted multi-user \ac{thz}-\ac{noma} system under beam misalignment by means of ergodic rate, \ac{op}, and \ac{do} under both discrete and random phase shifting configurations.
The authors in \cite{M12} optimized \ac{irs}-assisted sub-\ac{thz} systems under practical design constraints, evaluating received \ac{snr} and network throughput using low-complexity heuristic and Newton-Raphson-based algorithms.
The authors in \cite{M13} derived closed-form and asymptotic \ac{op} expressions for an \ac{irs}-assisted \ac{thz} system under cascaded $\alpha-\mu$ fading.
The authors in \cite{M14} analyzed an \ac{irs}-assisted \ac{thz} \ac{siso} system under $\alpha-\mu$ fading and pointing errors, deriving upper bound expressions for \ac{acc} and approximate \ac{op}.
The authors in \cite{M15} proposed a polar-domain channel estimation scheme for near-field \ac{irs}-assisted wideband \ac{thz} systems, evaluating normalized mean square error performance under frequency-wideband and spherical wavefront effects using block-sparse recovery.
In \cite{HINDUSTANI_IRS}, the authors analyzed the \ac{op} of a multi-\ac{irs}-assisted wireless system over Nakagami-$m$ fading channels by employing the \ac{clt} and \ac{lse} approach.
In \cite{wani2025exactsumdistributionalphaetakappamu}, the authors presented exact analytical frameworks employing multivariate Fox-H functions to evaluate \ac{op} and \ac{aber} of \ac{irs}-assisted wireless systems under generalized $\alpha-\eta-\kappa-\mu$ fading channels.
In \cite{IRS_Quantizer_Paper}, the authors observed that the number of active RF chains can be reduced by increasing the number of passive IRS elements. They also demonstrated that a 3-bit quantized IRS is sufficient to achieve good performance in terms of outage probability and wireless power transfer efficiency. 
A detailed comparison of the existing literature is provided in Table~\ref{SOTA_Table}, which outlines the state-of-the-art developments relevant to the proposed study.
\begin{table*}[t!]
  \centering
  \caption{State-of-the-art on IRS-assisted THz Systems}
  \label{SOTA_Table}
  \resizebox{\textwidth}{!}{%
\begin{tabular}{|c|c|c|c|c|cc|c|c|c|c|c|}
\hline
\multirow{2}{*}{\textbf{Ref.}} 
& \multirow{2}{*}{\textbf{THz Link}} 
& \multirow{2}{*}{\textbf{\makecell{Pointing\\Error}}} 
& \multirow{2}{*}{\textbf{IRS}} 
& \multirow{2}{*}{\textbf{DNN}} 
& \multicolumn{2}{c|}{\textbf{Modulation Scheme}} 
& \multirow{2}{*}{\textbf{OP}} 
& \multirow{2}{*}{\textbf{ACC}} 
& \multirow{2}{*}{\textbf{ASER}} 
& \multirow{2}{*}{\textbf{\makecell{Diversity\\Gain}}} 
& \multirow{2}{*}{\textbf{\makecell{Coding\\Gain}}} \\ \cline{6-7}
& & & & & \multicolumn{1}{c|}{\textbf{Binary Schemes}} & \textbf{M-ary Schemes} & & & & & \\ \hline

\cite{boulogeorgos2020outage}  & \checkmark & \checkmark & \ding{55} & \ding{55} & \multicolumn{1}{c|}{\ding{55}} & \ding{55} & \checkmark & \ding{55} & \ding{55} & \ding{55} & \ding{55} \\ \hline
\cite{M1}                      & \checkmark & \ding{55}  & \checkmark & \ding{55} & \multicolumn{1}{c|}{\ding{55}} & \ding{55} & \checkmark & \checkmark & \ding{55} & \ding{55} & \ding{55} \\ \hline
\cite{M2}                      & \checkmark & \ding{55}  & \checkmark & \ding{55} & \multicolumn{1}{c|}{\ding{55}} & \ding{55} & \ding{55} & \checkmark & \ding{55} & \ding{55} & \ding{55} \\ \hline
\cite{M3}                      & \checkmark & \ding{55}  & \checkmark & \ding{55} & \multicolumn{1}{c|}{\ding{55}} & \ding{55} & \ding{55} & \ding{55} & \ding{55} & \ding{55} & \ding{55} \\ \hline
\cite{M4}                      & \checkmark & \ding{55}  & \checkmark & \ding{55} & \multicolumn{1}{c|}{\ding{55}} & \ding{55} & \ding{55} & \checkmark & \ding{55} & \ding{55} & \ding{55} \\ \hline
\cite{M5}                      & \checkmark & \ding{55}  & \checkmark & \ding{55} & \multicolumn{1}{c|}{\begin{tabular}[c]{@{}c@{}}BPSK, BFSK,\\ DBPSK\end{tabular}} & QPSK, MQAM & \checkmark & \checkmark & \checkmark & \ding{55} & \ding{55} \\ \hline
\cite{M6}                      & \checkmark & \ding{55}  & \checkmark & \ding{55} & \multicolumn{1}{c|}{BPSK} & M-PSK & \checkmark & \checkmark & \checkmark & \checkmark & \ding{55} \\ \hline
\cite{M7}                      & \checkmark & \ding{55}  & \checkmark & \ding{55} & \multicolumn{1}{c|}{\ding{55}} & \ding{55} & \ding{55} & \checkmark & \ding{55} & \ding{55} & \ding{55} \\ \hline
\cite{M8}                      & \checkmark & \ding{55}  & \checkmark & \ding{55} & \multicolumn{1}{c|}{\ding{55}} & \ding{55} & \ding{55} & \checkmark & \ding{55} & \ding{55} & \ding{55} \\ \hline
\cite{M9}                      & \checkmark & \ding{55}  & \checkmark & \ding{55} & \multicolumn{1}{c|}{\ding{55}} & \ding{55} & \ding{55} & \checkmark & \ding{55} & \ding{55} & \ding{55} \\ \hline
\cite{M10}                     & \checkmark & \ding{55}  & \checkmark & \ding{55} & \multicolumn{1}{c|}{\ding{55}} & \ding{55} & \ding{55} & \checkmark & \ding{55} & \ding{55} & \ding{55} \\ \hline
\cite{M11}                     & \checkmark & \checkmark & \checkmark & \ding{55} & \multicolumn{1}{c|}{\ding{55}} & \ding{55} & \checkmark & \checkmark & \ding{55} & \checkmark & \ding{55} \\ \hline
\cite{M12}                     & \checkmark & \ding{55}  & \checkmark & \ding{55} & \multicolumn{1}{c|}{\ding{55}} & \ding{55} & \ding{55} & \checkmark & \ding{55} & \ding{55} & \ding{55} \\ \hline
\cite{M13}                     & \checkmark & \ding{55}  & \checkmark & \ding{55} & \multicolumn{1}{c|}{\ding{55}} & \ding{55} & \checkmark & \ding{55} & \ding{55} & \ding{55} & \ding{55} \\ \hline
\cite{M14}                     & \checkmark & \checkmark & \checkmark & \ding{55} & \multicolumn{1}{c|}{\ding{55}} & \ding{55} & \checkmark & \checkmark & \ding{55} & \ding{55} & \ding{55} \\ \hline
\cite{M15}                     & \checkmark & \ding{55}  & \checkmark & \ding{55} & \multicolumn{1}{c|}{\ding{55}} & \ding{55} & \ding{55} & \ding{55} & \ding{55} & \ding{55} & \ding{55} \\ \hline
\cite{HINDUSTANI_IRS}          & \ding{55}  & \ding{55}  & \checkmark & \ding{55} & \multicolumn{1}{c|}{\ding{55}} & \ding{55} & \checkmark & \ding{55} & \ding{55} & \checkmark & \checkmark \\ \hline
\textbf{[This work]}           & \checkmark & \checkmark & \checkmark & \checkmark & \multicolumn{1}{c|}{\begin{tabular}[c]{@{}c@{}}BPSK, BFSK,\\ DBPSK\end{tabular}} & \begin{tabular}[c]{@{}c@{}}MQAM, RQAM, \\ HQAM\end{tabular} & \checkmark & \checkmark & \checkmark & \checkmark & \checkmark \\ \hline
\end{tabular}%
  } 
\end{table*}
\subsection{Research Gap and Motivation} \label{Research_gap_and_motivation}
Despite these efforts, accurate and tractable \ac{irs}-assisted \ac{thz} channel modelling remains challenging due to phase errors, beam misalignment, and complex stochastic fading behaviours. Existing approaches based on the \ac{clt} provide approximate results that are valid mainly for large \ac{irs} element counts. The \ac{lse} technique, however, offers a more accurate \ac{e2e} statistical representation even for a moderate number of reflective elements, yet its integration with \ac{thz} channel impairments such as asymmetric $\alpha-\mu$ fading and pointing errors is largely unexplored.
\subsection{Contributions}
Motivated by the above observations, this paper presents a hybrid analytical–learning framework for \ac{irs}-assisted \ac{thz} communication systems. The key contributions are summarized as follows:
\begin{enumerate}
    \item \textbf{Analytical Modeling:} Development of a unified analytical model for \ac{irs}-assisted \ac{thz} links incorporating atmospheric attenuation, asymmetric $\alpha-\mu$ small-scale fading, and pointing errors, with the \ac{lse} approach used to approximate the composite IRS channel.
    \item \textbf{Closed-Form Expressions:} Derivation of closed-form \ac{op}, \ac{acc}, and \ac{aser} expressions for \ac{rqam} and \ac{hqam} modulations, including asymptotic results for \ac{do} and \ac{cg}.
    \item \textbf{\ac{dnn}-Aided Prediction:} Design of lightweight \ac{dnn} frameworks to predict \ac{op} and \ac{aser} with minimal computational cost, providing near-instantaneous estimation accuracy comparable to analytical solutions.
    \item \textbf{Comprehensive Validation:} Monte Carlo simulations are conducted to validate analytical results and to investigate the impact of \ac{irs} element count, phase quantization, and random co-phasing errors under practical conditions.
\end{enumerate}
\subsection{Organization of the Paper}\label{Organization_of_the_paper}
The rest of the paper is organized as follows. In Section~\ref{System and Channel Model}, the system and channel model are presented, including the modeling of atmospheric attenuation, asymmetrical $\alpha\!-\!\mu$ small-scale fading, and pointing errors. In Section~\ref{Analytical Performance analysis}, closed-form analytical expressions for the \ac{op}, \ac{acc}, and \ac{aser} for \ac{rqam} and \ac{hqam} are derived, and the asymptotic analyses are provided to characterize the coding gain and diversity order. In Section~\ref{DNN Section}, the \ac{dnn}-based frameworks are developed for rapid prediction of \ac{op} and \ac{aser}, and the network architectures and training methodology are described. In Section~\ref{Numerical and simulation results}, the numerical results and Monte Carlo simulations are presented to validate the analysis and illustrate the impact of system parameters, \ac{irs} phase errors, and element count. Finally, in Section~\ref{conclusion}, the paper is concluded. The detailed integral solutions used in Section~\ref{Analytical Performance analysis} are provided in Appendices~\ref{Appendix_I1}--\ref{Appendix_I5}.

\subsection{Notations} \label{notations}
This subsection summarizes the mathematical symbols and functions employed throughout the manuscript. The lower and upper incomplete gamma functions \cite[eqs. (8.350.1)–(8.350.2)]{gradshteyn2014table} are denoted by $\gamma(\cdot,\cdot)$ and $\Gamma(\cdot,\cdot)$, respectively, while $\Gamma(\cdot)$ represents the complete gamma function \cite[eq. (8.310.1)]{gradshteyn2014table}. The confluent hypergeometric function of the first kind \cite[eq. (9.210.1)]{gradshteyn2014table} is expressed as ${}_{1}F_{1}(\cdot;\cdot;\cdot)$, and the generalized hypergeometric function \cite[eq. (7.2.3)]{prudnikov} is written as ${}_{p}F_{q}(a_{1},\ldots,a_{p}; b_{1},\ldots,b_{q}; z)$. The Meijer $G$-function \cite[eq. (9.301)]{gradshteyn2014table} is represented by $G_{p, q}^{m, n}\!\left[z \,\middle|\, \begin{array}{c} a_{1}, \ldots, a_{p} \\ b_{1}, \ldots, b_{q} \end{array}\right]$, while the Fox $H$-function \cite[eq. (17)]{SRIVASTAVA1979191} is denoted as $H_{p, q}^{m, n}\!\left[z \,\middle|\, \begin{array}{c} (a_{1}, A_{1}), \ldots, (a_{p}, A_{p}) \\ (b_{1}, B_{1}), \ldots, (b_{q}, B_{q}) \end{array}\right]$. Furthermore, the multivariate Fox $H$-function \cite[eq. (A.1)]{Mathai2010} is given by
\begin{align}
\resizebox{\columnwidth}{!}{$
H_{p,q:p_1,q_1;\ldots;p_r,q_r}^{0,n:m_1,n_1;\ldots;m_r,n_r}
\!\left[
\begin{array}{c}
z_1 \\[-1pt]
\vdots \\[-1pt]
z_r
\end{array}
\middle|
\begin{array}{c}
(a;\alpha_1,\ldots,\alpha_r)_{1:p}:(c,\gamma)_{1:p_1};\ldots;(c,\gamma)_{1:p_r} \\
(b;\beta_1,\ldots,\beta_r)_{1:q}:(d,\delta)_{1:q_1};\ldots;(d,\delta)_{1:q_r}
\end{array}
\right]
$}
\nonumber.
\end{align}
\section{System and Channel Model}\label{System and Channel Model}
\begin{figure}[t!]
    \centering
    \includegraphics[width=8cm,height=4cm]{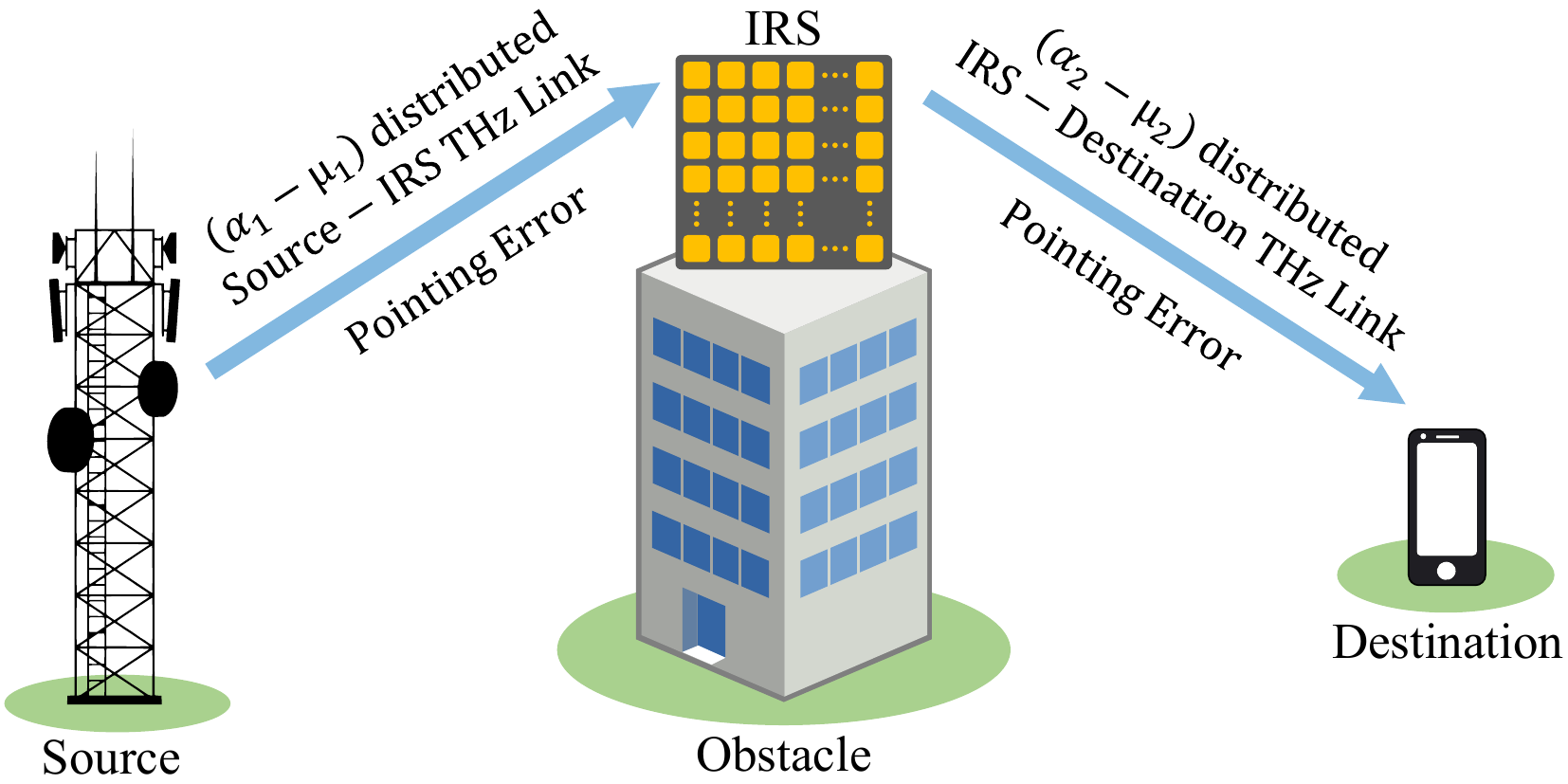}
    \caption{IRS-assisted asymmetrical THz system model with pointing errors}
    \label{System_model_Fig}
\end{figure}

We adopt a \ac{thz} link between a \ac{s} and a \ac{d}, where direct transmission is obstructed due to geographical or architectural constraints. To establish an indirect link, an $N$ element \ac{irs} is strategically deployed between them, as depicted in Fig. \ref{System_model_Fig}. 
However, both links are affected by \ac{fspl}, atmospheric attenuation, $\alpha-\mu$ distributed small-scale fading and pointing errors, which degrade overall system performance. 
\subsection{System Model}
We consider a downlink transmission where a signal with power $P_s$ is transmitted from \ac{s} and reaches \ac{d} after being reflected by the \ac{irs}. The received signal at \ac{d} is expressed as
\begin{align}\label{received_Signal}
    y_d = \sqrt{P_s} h_a \sum_{j=1}^N h_{1j}h_{2j} + \omega_d,
\end{align}
where the molecular absorption coefficient $\left(h_a\right)$ is discussed in Subsection~\ref{Atmospheric_Attenuation}, whereas $h_{ij},~i \in \{1,2\},~j \in [1,N]$, represents the channel coefficient of $i^{th}$ hop associated with the $j^{\text{th}}$ \ac{irs} element. Specifically, $i=1$ corresponds to the incident link from \ac{s} to the $j^{\text{th}}$ \ac{irs} element, and $i=2$ corresponds to the reflected link from the $j^{\text{th}}$ \ac{irs} element to \ac{d}. The term $\omega_d$ denotes the received \ac{awgn} at \ac{d}, modeled as a zero-mean complex Gaussian random variable with variance $\sigma_n^2$. Further, by considering $h_{ij} = h_{l_{ij}}h_{f_{ij}}$, eq. \eqref{received_Signal} can be reformulated as
\begin{align}\label{received_Signal_2}
    y_d = \sqrt{P_s} h_a \sum_{j=1}^N \prod_{i=1}^2 h_{l_{ij}}h_{f_{ij}} + \omega_d.
\end{align}
The term $h_{f_{ij}}$ represents the combined effect of small-scale fading and pointing error in the $i^{\text{th}}$ hop, with its stochastic behavior detailed in Subsection~\ref{THz_link_Channel_Model}. On the other hand, $h_{l_{ij}}$ denotes the large-scale fading coefficient for the $i^{\text{th}}$ hop and can be mathematically expressed as
\begin{align} \label{FSPL_Expression}
    h_{l_{ij}} = \frac{c \sqrt{G_{t_i}G_{r_i}}}{4\pi f}\, d_{ij}^{-0.5\eta},
\end{align}
where $f$ denotes the \ac{thz} carrier frequency and $c$ is the speed of light in free space. The antenna gains at the transmitter and receiver are represented by $G_{t_1}$ and $G_{r_2}$, respectively. Unity gains are assumed at the \ac{irs}, i.e., $G_{t_2}=G_{r_1}=1$, since only the phase shift due to reflection is considered. The path-loss exponent is set to $\eta=2$ without shadowing, although in dense metropolitan environments, higher values of $\eta$ may be observed.
\subsection{Atmospheric Attenuation} \label{Atmospheric_Attenuation}
The resonance frequencies of water vapor molecules fall within the \ac{thz} band. Consequently, when signals propagate at \ac{thz} frequencies, the water vapor molecules in the atmosphere resonate, leading to molecular absorption. The deterministic molecular absorption coefficient $\left(h_a\right)$ depends on environmental factors such as absolute temperature $\left(T\right)$, atmospheric pressure $\left(P\right)$, and relative humidity $\left(\psi\right)$. Mathematically, $h_a$ can be expressed using Buck's equation \cite[eq. (4)]{Soum_PC} as
\begin{align}
    h_a = \exp\!\left[-0.5\,k_\alpha\!\left(f, T, P, \psi\right)\right],
\end{align}
where the absorption coefficient $k_\alpha(f, T, P, \Psi)$, which depends on frequency and environmental factors, is modeled as given in \cite[eq.~(8)]{boulogeorgos2019analytical}.
\subsection{Small-Scale Fading} \label{THz_link_Channel_Model} 
We consider all the \ac{s}-to-\ac{irs} and \ac{irs}-to-\ac{d} links to be jointly affected by asymmetrical pointing errors $(h_{p_{ij}})$ under \ac{inid} $\alpha$–$\mu$ distributed small-scale fading $(h_{sc_{ij}})$ in our analysis. Therefore, the \ac{cdf} of $h_{f_{ij}}$ can be obtained with the aid of \cite[eq.~(3)]{Soum_WCL} as
\begin{align}\label{Individual_Link_CDF_Internal_step_1}
    F_{h_{f_{ij}}}\left(\chi\right) = \int_{y=0}^{S_{0_{ij}}}F_{h_{sc_{ij}}}\left(\frac{\chi}{y}\right)f_{h_{p_{ij}}}\left(y\right)dy,
\end{align}
where the \ac{pdf} of $h_{p_{ij}}$ and the \ac{cdf} of $h_{sc_{ij}}$ can be expressed with the aid of \cite[eqs.~(1)~and~(2)]{Soum_WCL}, as given in eq. \eqref{PDF_h_a} and eq. \eqref{CDF_h_sc}, respectively.
\begin{align}
    f_{h_{p_{ij}}} \left(\chi\right) &= \frac{\phi_{ij}}{S_{0_{ij}}^{\phi_{ij}}}\chi^{\phi_{ij}-1}, \quad 0 \leq \chi \leq S_{0_{ij}}, \label{PDF_h_a} \\
    F_{h_{sc_{ij}}}\left(\chi\right) &= \frac{1}{\Gamma\left(\mu_{ij}\right)}\gamma\left[\mu_{ij},~\mu_{ij}\left(\frac{\chi}{\Omega_{ij}}\right)^{\alpha_{ij}}\right], \quad \chi>0 \label{CDF_h_sc}
\end{align}
where $\alpha_{ij}$, $\mu_{ij}$, and $\Omega_{ij}$ denote the small-scale fading parameters, while $\phi_{ij}$ and $S_{0_{ij}}$ represent the pointing error coefficients.
Accordingly, substituting eq. \eqref{PDF_h_a} and eq. \eqref{CDF_h_sc} in eq. \eqref{Individual_Link_CDF_Internal_step_1}, the \ac{cdf} of $h_{f_{ij}}$ can be derived similar to \cite[eq.~(6)]{Soum_WCL} as
\begin{align}\label{Individual_Link_CDF}
    F_{h_{f_{ij}}}\left(\chi\right) = 1 - \frac{\phi_{ij}}{\alpha_{ij} \Gamma\left(\mu_{ij}\right)}G_{2,3}^{3,0}\left[\frac{\mu_{ij}\chi^{\alpha_{ij}}}{\Omega_{ij}^{\alpha_{ij}}S_{0_{ij}}^{\alpha_{ij}}}\left|\begin{array}{cc}
         1,\frac{\phi_{ij}}{\alpha_{ij}}+1  \\
         \mu_{ij}, 0, \frac{\phi_{ij}}{\alpha_{ij}}
    \end{array}\right.\right], 
\end{align}
Furthermore, the \ac{pdf} of ${h_f}_{ij}$ is obtained by differentiating eq. (\ref{Individual_Link_CDF}) with respect to $\chi$ by using the integral form representation of the Meijer-$G$ function given in \refMG,  as
\begin{align}\label{Individual_link_pdf_intermediate_1}
    f_{h_{f_{ij}}}\left(\chi\right) = \frac{1}{2\pi i}\int_s&\frac{\phi_{ij}\Gamma\left(\mu_{ij}+s\right)\Gamma\left(\frac{\phi_{ij}}{\alpha_{ij}}+s\right)}{\chi\Gamma\left(\mu_{ij}\right)\Gamma\left(\frac{\phi_{ij}}{\alpha_{ij}}+1+s\right)} \left[\frac{\Omega_{ij}^{\alpha_{ij}}S_{0_{ij}}^{\alpha_{ij}}}{\mu_{ij}\chi^{\alpha_{ij}}}\right]^{s}ds.
\end{align}
Again, using the representation in \refMG, eq. (\ref{Individual_link_pdf_intermediate_1}) can be expressed in terms of Meijer-$G$ function as
\begin{align}\label{IRS_individual_link_pdf}
    f_{h_{f_{ij}}}\left(\chi\right) = \frac{\phi_{ij}}{\chi\Gamma\left(\mu_{ij}\right)}G_{1,2}^{2,0}\left[\frac{\mu_{ij}\chi^{\alpha_{ij}}}{\Omega_{ij}^{\alpha_{ij}}S_{0_{ij}}^{\alpha_{ij}}}\left|\begin{array}{cc}
         \frac{\phi_{ij}}{\alpha_{ij}}+1  \\
         \mu_{ij}, \frac{\phi_{ij}}{\alpha_{ij}}
    \end{array}\right.\right].
\end{align}
Furthermore, the $n^{\text{th}}$ moment of the $i^{\text{th}}$ hop in the considered system can be obtained as
\begin{align}\label{nth_moment_individual_link}
    \mathbb{E}\left[\chi^n\right]=\int_{\chi=-\infty}^\infty\chi^n f_{h_{f_{ij}}}\left(\chi\right)d\chi.
\end{align}
By substituting eq. (\ref{IRS_individual_link_pdf}) into eq. (\ref{nth_moment_individual_link}), it can be rewritten as
\begin{align}\label{nth_moment_individual_link_intermediate}
    \mathbb{E}\left[\chi^n\right]=\frac{\phi_{ij}}{\Gamma\left(\mu_{ij}\right)}\int_{\chi=0}^\infty\chi^{n-1}G_{1,2}^{2,0}\left[\frac{\mu_{ij}\chi^{\alpha_{ij}}}{\Omega_{ij}^{\alpha_{ij}}S_{0_{ij}}^{\alpha_{ij}}}\left|\begin{array}{cc}
         \frac{\phi_{ij}}{\alpha_{ij}}+1  \\
         \mu_{ij}, \frac{\phi_{ij}}{\alpha_{ij}}    
    \end{array}\right.\right]d\chi.
\end{align}
By applying the change of variable $\chi^\alpha = t$, eq. (\ref{nth_moment_individual_link_intermediate}) can be reformulated as
\begin{align}\label{nth_moment_individual_link_intermediate_2}
    \mathbb{E}\left[\chi^n\right]=\frac{\phi_{ij}}{\alpha_{ij}\Gamma\left(\mu_{ij}\right)}\int_{t=0}^\infty t^{\frac{n}{\alpha_{ij}}-1}G_{1,2}^{2,0}\left[\frac{\mu_{ij}t}{\Omega_{ij}^{\alpha_{ij}}S_{0_{ij}}^{\alpha_{ij}}}\left|\begin{array}{cc}
         \frac{\phi_{ij}}{\alpha_{ij}}+1  \\
         \mu_{ij}, \frac{\phi_{ij}}{\alpha_{ij}}
    \end{array}\right.\right]dt.
\end{align}
Finally, with the aid of \cite[eq. (07.34.21.0009.01)]{wolfram_functions}, eq. (\ref{nth_moment_individual_link_intermediate_2}) can be simplified as
\begin{align}\label{IRS_individual_link_nth_moment}
    \mathbb{E}\left[\chi^n\right] = \frac{\phi_{ij}\Omega^n_{ij}S_{0_{ij}}^n}{\mu_{ij}^{\frac{n}{\alpha_{ij}}}\left(\phi_{ij}+n\right)}\frac{\Gamma\left(\frac{n}{\alpha_{ij}}+\mu_{ij}\right)}{\Gamma\left(\mu_{ij}\right)}.
\end{align}
Now, by substituting $n=1$ and $n=2$ into eq. (\ref{IRS_individual_link_nth_moment}), the first and second moments, $\mathbb{E}\!\left[\chi\right]$ and $\mathbb{E}\!\left[\chi^{2}\right]$, can be obtained. Furthermore, the mean ($\mu_{ij}$) and variance ($\sigma_{ij}^{2}$) of $h_{f_{ij}}$ can be derived using the relations $\mu_{ij} = \mathbb{E}\!\left[\chi\right]$ and $\sigma_{ij}^{2} = \mathbb{E}\!\left[\chi^{2}\right] - \left(\mathbb{E}\!\left[\chi\right]\right)^{2}$, respectively, as
\begin{align}
    \mu_{ij} &= \frac{\phi_{ij}\Omega_{ij} S_{0_{ij}}}{\mu_{ij}^{\frac{1}{\alpha_{ij}}}\left(\phi_{ij}+1\right)}\frac{\Gamma\left(\frac{1}{\alpha_{ij}}+\mu_{ij}\right)}{\Gamma\left(\mu_{ij}\right)}, \\
    \sigma_{ij}^2 &= \frac{\phi_{ij}S_{0_{ij}}^2\Omega_{ij}^2}{\mu_{ij}^{\frac{2}{\alpha_{ij}}}\Gamma\left(\mu_{ij}\right)}\frac{\Gamma\left(\frac{2}{\alpha_{ij}}+\mu_{ij}\right)}{\phi_{ij}+2}-\mu^2_{ij}.
\end{align}
By considering $h_{f_{j}}=\prod_{i=1}^2 h_{f_{ij}} = h_{f_{1j}} h_{f_{2j}}$, the mean and variance of $h_{f_{j}}$ are given by $\mu_{h_{j}} = \mu_{1j}\mu_{2j}$ and $\sigma_{h_j}^2 = \sigma_{1j}^2\sigma_{2j}^2 + \sigma_{1j}^2\mu_{2j}^2 + \mu_{1j}^2\sigma_{2j}^2$, respectively, as expressed in eqs. (\ref{IRS_Mean}) and (\ref{IRS_Variance}).
\begin{align}
    \mu_{h_j} =& \prod_{i=1}^2\frac{\phi_{ij}S_{0_{ij}}\Omega_{ij}}{\phi_{ij}+1}\frac{\Gamma\left(\frac{1}{\alpha_{ij}}+\mu_{ij}\right)}{\mu_{ij}^{\frac{1}{\alpha_{ij}}}\Gamma\left(\mu_{ij}\right)} \label{IRS_Mean}, \\
    \sigma_{h_j}^2=&\prod_{i=1}^2\frac{\phi_{ij}S_{0_{ij}}^2\Omega_{ij}^2}{ \mu_{ij}^{\frac{2}{\alpha_{ij}}} 
    }\frac{\Gamma\left(\frac{2}{\alpha_{ij}}+\mu_{ij}\right)}{\left(\phi_{ij}+2\right)\Gamma\left(\mu_{ij}\right)} \nonumber
    \\ &-\prod_{i=1}^2\frac{\phi_{ij}^2S_{0_{ij}}^2\Omega_{ij}^2}{ \mu_{ij}^{\frac{2}{\alpha_{ij}}}}\left[\frac{\Gamma\left(\frac{1}{\alpha_{ij}}+\mu_{ij}\right)}{ \left(\phi_{ij}+1\right)  
   \Gamma\left(\mu_{ij}\right)}\right]^2. \label{IRS_Variance}
\end{align}

\subsection{\texorpdfstring{\ac{cdf} of \ac{e2e} \ac{snr}}{CDF of E2E SNR}}\label{sub_cdf}
The \ac{e2e} \ac{snr} ($\lambda$) in the considered system can be obtained from eq. (\ref{received_Signal_2}) as
\begin{align}\label{snr_e2e}
    \lambda=\left| h_{a} \sum_{j=1}^N \prod_{i=1}^{2} h_{l_{ij}} h_{f_{ij}}\right|^2 \frac{P_s}{\sigma_n^2}.
\end{align}
Here, the average transmit \ac{snr} $\left(\bar{\lambda}_{0}\right)$ can be defined as $\bar{\lambda}_{0} = \frac{P_s}{\sigma_{n}^2}$. Before deriving the \ac{cdf} of $\lambda$, it is necessary to obtain the \ac{pdf} and \ac{cdf} of $\mathcal{B} = \sum_{j=1}^N\prod_{i=1}^2 h_{f_{ij}}$, which can be computed using the \ac{lse} method \cite[eqs. (12) and (13)]{HINDUSTANI_IRS} as
\begin{align}
    f_{\mathcal{B}}\left(b\right) &= \frac{b^\tau e^-\frac{b}{\Lambda}}{\Lambda^{\tau+1}\Gamma\left(\tau+1\right)}, \\
    F_{\mathcal{B}}\left(b\right) &= \frac{1}{\Gamma\left(\tau+1\right)}\gamma\left(\tau+1,\frac{b}{\Lambda}\right),
\end{align}
where $\Lambda = \frac{\sigma_{\mathcal{B}}^2}{\mu_{\mathcal{B}}}$ and $\tau = \frac{\mu_{\mathcal{B}}^2}{\sigma_{\mathcal{B}}^2} - 1$, while the mean and variance of $\mathcal{B}$ are expressed in terms of $\mu_{h_{j}}$ and $\sigma^{2}_{h_{j}}$ as
$\begin{aligned}
    \mu_\mathcal{B} = \sum_{j=1}^N \mu_{h_j}, \text{ and } \sigma_\mathcal{B}^2 =  \sum_{j=1}^N \sigma_{h_j}^2.
\end{aligned}$
Accordingly, the \ac{cdf} for $\lambda$ at \ac{d} can be expressed as
\begin{align}\label{CDF_Expression}
    F_\lambda\left(\lambda\right) = F_\mathcal{B}\left(\sqrt{\frac{\lambda}{\lambda_0}}\right) = \frac{1}{\Gamma\left(\tau+1\right)}\gamma\left(\tau+1,\frac{\sqrt{\lambda}}{\Lambda\sqrt{\lambda_0}}\right),
\end{align}
where, $\lambda_0 = \left|h_a \sum_{j=1}^N\prod_{i=1}^2 h_{l_{ij} h_{f_{ij}}}\right|^2$.
The statistical characterization of the \ac{e2e} \ac{snr} obtained here serves as the foundation for the analytical derivations of \ac{op}, \ac{aser}, and \ac{acc} in Section III.
\section{Analytical Performance Analysis}\label{Analytical Performance analysis}
This section quantifies the reliability and efficiency of the \ac{irs}-assisted \ac{thz} link through three key metrics: \ac{op}, \ac{aser}, and \ac{acc}. \ac{op} indicates the probability of link failure below an \ac{snr} threshold, \ac{aser} measures modulation accuracy, and \ac{acc} reflects achievable information rate under fading and pointing conditions.
\subsection{Outage Probability}
The \ac{op} is defined as the probability that the received \ac{snr} falls below threshold \ac{snr} $\left(\lambda_{th}\right)$. Mathematically, \ac{op} can be expressed as $P_O\left(\lambda_{th}\right)=P_r\left(\lambda<\lambda_{th}\right)=F_\lambda\left(\lambda_{th}\right)$. Therefore, \ac{op} can be expressed with the aid of eq. (\ref{CDF_Expression}) as
\begin{align}\label{Outage_Probability}
    P_O\left(\lambda_{th}\right) = \frac{1}{\Gamma\left(\tau+1\right)}\gamma\left(\tau+1, \frac{\sqrt{\lambda_{th}}}{\Lambda\sqrt{\lambda_0}}\right).
\end{align}
The closed-form \ac{op} expression derived above enables subsequent evaluation of the \ac{aser} for various \ac{qam} formats.

\textbf{Remark 1:} 
The closed-form \ac{op} expression in~(\ref{Outage_Probability}), represented in terms of the lower incomplete Gamma function 
$\gamma(\cdot,\cdot)$, reveals the cumulative behavior of the received \ac{snr} below the threshold $\lambda_{\mathrm{th}}$. 
The incomplete-Gamma form indicates that OP decreases sharply with the normalized \ac{snr} parameter 
$\sqrt{\lambda_{\mathrm{th}}/\Lambda^2\lambda_{0}}$, which is governed by the fading parameters $\alpha$ and $\mu$ 
as well as the effective IRS gain $\Lambda$. 
Larger values of $\alpha$ correspond to weaker amplitude nonlinearity, whereas higher $\mu$ denotes a greater number of 
multipath clusters—both leading to a smoother channel envelope and hence a smaller argument of 
$\gamma(\cdot,\cdot)$, resulting in reduced outage probability. 
Conversely, an increase in pointing-error variance or a reduction in the number of IRS elements diminishes $\Lambda$, 
thereby enlarging the incomplete-Gamma argument and worsening the \ac{op} performance. 
This analytical behavior agrees with the simulation trends, confirming that higher $\alpha$ and $\mu$ 
enhance link reliability by mitigating fading severity. 
%
%
%
\subsection{Average Symbol Error Rate}
With the aid of \ac{cdf}-based approach, the \ac{aser} for any coherent modulation scheme can be obtained as
\begin{align}\label{ASER_Definition}
    P_s\left(e\right) = -\int_{\lambda=0}^{\infty}P_{s}^{'}\left(e|\lambda\right)F_\lambda\left(\lambda\right) \ d\lambda,
\end{align}
where $P_{s}^{'}\left(e|\lambda\right)$ denotes the first-order derivative of the conditional \ac{aser} with respect to $\lambda$ for the corresponding modulation scheme.
\subsubsection{\texorpdfstring{Coherent \ac{rqam} scheme}{Coherent RQAM scheme}}\label{Subsubsection_Coherent_RQAM}
Using \cite[eq. (4)]{Nagendra_Sir_RQAM_Cond_ASER_EQ4}, the first-order derivative of the conditional \ac{aser} for the $M_I \times M_Q$ \ac{rqam} scheme can be expressed as
\begin{align}\label{RQAM_Cond_ASER}
    P_{s}^{'R}\left(e|\lambda\right) =& \frac{1}{\sqrt{\lambda}}\left[\mathcal{D}e^{-\left(0.5\lambda a^2\right)}+\mathcal{F}e^{-\left(0.5 \lambda b^2\right) }\right] \nonumber \\ - \frac{\mathcal{G}}{\sqrt{\pi}}
    &e^{-\frac{\lambda}{2}\left(a^2+b^2\right)}\left[{}_{1}F_{1}\left(1;\frac{3}{2};\frac{\lambda a^2}{2}\right) \right.
    +\left.{}_{1}F_{1}\left(1;\frac{3}{2};\frac{\lambda b^2}{2}\right)\right],
\end{align}
where $\mathcal{D}=\frac{ap\left(q-1\right)}{\sqrt{2\pi}}, \mathcal{F}=\frac{b\left(p-1\right)q}{\sqrt{2\pi}}$, $\mathcal{G}=\frac{abpq}{\pi}$ wherein, $a=\sqrt{\frac{6}{\left(M_I^2-1\right)^2+\left(M^2_Q-1\right)\beta^2}}$, $b=\beta a$, $p=1-\frac{1}{M_I}$, and $q=1-\frac{1}{M_Q}$. $M_I$ and $M_Q$ are the in-phase and quadrature-phase components, respectively. Furthermore, by substituting eqs. (\ref{CDF_Expression}) and (\ref{RQAM_Cond_ASER}) into eq. (\ref{ASER_Definition}), the \ac{aser} expression for \ac{rqam} can be obtained in terms of integral functions as 
\begin{align}\label{rqam_aser_intermediate_1}
    &P^{R}_{s}\left(e\right)=
    -\mathcal{D}\mathcal{I}_{1}\left(-0.5, 0.5a^2\right)
    -\mathcal{F}\mathcal{I}_{1}\left(-0.5, 0.5b^2\right) \nonumber \\
    &+\mathcal{G}\mathcal{I}_{3}\left(0, \frac{a^2+b^2}{2}, \frac{a^2}{2}\right)
    +\mathcal{G}\mathcal{I}_{3}\left(0, \frac{a^2+b^2}{2}, \frac{b^2}{2}\right) \nonumber \\
    &+\frac{\mathcal{D}}{\Gamma\left(\tau+1\right)}\mathcal{I}_{2}\left(-0.5, 0.5a^2\right)
    + \frac{\mathcal{F}}{\Gamma\left(\tau+1\right)}\mathcal{I}_{2}\left(-0.5, 0.5b^2\right) \nonumber \\
    &-\frac{\mathcal{G}}{\Gamma\left(\tau+1\right)}\left[\mathcal{I}_{4}\left(0, \frac{a^2+b^2}{2}, \frac{a^2}{2}\right) 
    +\mathcal{I}_{4}\left(0,\frac{a^2+b^2}{2}, \frac{b^2}{2}\right)\right],
\end{align}
\\
where $\mathcal{I}_{1}\left(\cdot, \cdot\right)$, $\mathcal{I}_2\left(\cdot, \cdot\right)$, $\mathcal{I}_{3}\left(\cdot, \cdot, \cdot\right)$ and $\mathcal{I}_{4}\left(\cdot, \cdot, \cdot\right)$ can be expressed as
\begin{align}
    &\mathcal{I}_{1}\left(\chi_1, \chi_2\right) = \int_{0}^\infty \frac{\lambda^{\chi_1}}{e^{\chi_2 \lambda}}d\lambda. \label{I1_Def}\\
    &\mathcal{I}_{2}\left(\chi_{1}, \chi_2\right) = \int_{0}^\infty\frac{\lambda^{\chi_1}}{e^{\chi_2 \lambda}}\Gamma\left(\tau+1, \frac{\sqrt{\lambda}}{\Lambda\sqrt{\lambda_0}}\right) d\lambda. \label{I2_Def}\\
    &\mathcal{I}_{3}\left(\chi_1, \chi_2, \chi_3\right) = \int_{0}^\infty\frac{\lambda^{\chi_1}}{e^{\chi_2\lambda}}{}_{1}F_{1}\left(1;\frac{3}{2};\chi_3\lambda\right) d\lambda . \label{I3_Def}
\end{align}
\begin{align}
    &\mathcal{I}_{4}\left(\chi_1, \chi_2, \chi_3\right) = \int_{\lambda=0}^\infty\frac{\lambda^{\chi_1}}{e^{\chi_2\lambda}}{}_{1}F_{1}\left(1;\frac{3}{2};\lambda\chi_3\right)
    \Gamma\left(\tau+1, \frac{\sqrt{\lambda}}{\Lambda\sqrt{\lambda_0}}\right) d\lambda. \label{I4_Def}
\end{align}
The solutions of $\mathcal{I}_{1}\left(\cdot, \cdot\right)$, $\mathcal{I}_2\left(\cdot, \cdot\right)$, $\mathcal{I}_{3}\left(\cdot, \cdot, \cdot\right)$ and $\mathcal{I}_{4}\left(\cdot, \cdot, \cdot\right)$ are discussed in appendices \ref{Appendix_I1}, \ref{Appendix_I2}, \ref{Appendix_I3} and \ref{Appendix_I4}, respectively. By substituting these integral solutions into eq. (\ref{rqam_aser_intermediate_1}), the generalized \ac{aser} expression for the \ac{rqam} can be expressed as in eq. \eqref{rqam_aser_intermediate_2}.
\begin{figure*}[t]
\noindent
\begin{align}\label{rqam_aser_intermediate_2}
    & P^{R}_{s}\left(e\right) = p + q - 2pq 
    + \frac{2\mathcal{G}}{a^2 + b^2} \left[ {}_2F_1\left(1,1; \tfrac{3}{2}; \tfrac{a^2}{a^2 + b^2}\right)
    + {}_2F_1\left(1,1; \tfrac{3}{2}; \tfrac{b^2}{a^2 + b^2}\right) \right]
    +\frac{1}{\Gamma\left(\tau+1\right)}
    \left\{\frac{\mathcal{D}\sqrt{2}}{a} H_{2,2}^{2,1} \!\left[\frac{\sqrt{2}}{\Lambda a\sqrt{\lambda_0}} \Bigg| 
    \begin{array}{cc}
         \nu_0\!\left(-\tfrac{1}{2}\right), \nu_1 \\
         \nu_2, \nu_3
    \end{array}\!\right]\right. \nonumber \\
    + &\frac{\mathcal{F}\sqrt{2}}{b}
    H_{2,2}^{2,1} \!\left[\frac{\sqrt{2}}{\Lambda b\sqrt{\lambda_0}} \Bigg| 
    \begin{array}{cc}
         \nu_0\!\left(-\tfrac{1}{2}\right), \nu_1 \\
         \nu_2, \nu_3
    \end{array}\!\right] 
    - \frac{\mathcal{G}}{a^2} H_{1,0:1,2;1,2}^{0,1:1,1;2,0}\!\left[
    \begin{array}{cc}
            \tfrac{b^2}{a^2}  \\
            \tfrac{\sqrt{2}}{a\Lambda\sqrt{\lambda_0}}
    \end{array}\!\Bigg|
    \begin{array}{c}
        \nu_6(0):\nu_7;\nu_1 \\
        -:\nu_3,\nu_8;\nu_2,\nu_3
    \end{array}\!\right] 
    -\frac{\mathcal{G}}{b^2}
    H_{1,0:1,2;1,2}^{0,1:1,1;2,0}\!\left[
    \begin{array}{cc}
            \tfrac{a^2}{b^2}  \\
            \tfrac{\sqrt{2}}{b\Lambda\sqrt{\lambda_0}}
    \end{array}\!\Bigg|
    \begin{array}{c}
        \nu_6(0):\nu_7;\nu_1 \\
        -:\nu_3,\nu_8;\nu_2,\nu_3
    \end{array}\!\right].
\end{align}
\rule{\textwidth}{0.4pt}
\end{figure*}

\subsubsection{\texorpdfstring{Coherent \Ac{hqam} scheme}{Coherent HQAM scheme}}\label{Subsubsection_Coherent_HQAM}
With the aid of 
\cite[eq. (7)]{Nagendra_Sir_HQAM_Cond_ASER_Eq7}, the first order derivative of conditional \ac{aser} for $M-$\ac{hqam} scheme can be expressed as
\begin{align}\label{HQAM_Cond_ASER}
    P^{'H}_{s}\left(e|\lambda\right) &= 
    \sqrt{\frac{\alpha_h}{2\pi\lambda}}\left(\frac{B_c-B}{2}\right)e^{-\frac{\alpha_{h}\lambda}{2}} 
    - \sqrt{\frac{\alpha_h}{3\pi\lambda }}\left(\frac{B_c}{3}\right) \nonumber \\
    &\times e^{-\frac{\alpha_{h}\lambda}{3}} 
    + \sqrt{\frac{\alpha_h}{6\pi}}\left(\frac{B_c}{2}\right)\lambda^{-0.5}e^{-\frac{\alpha_{h}\lambda}{6}}
    - \frac{B_c \alpha_h}{2\sqrt{3}\pi} \nonumber \\
    &\times e^{-\frac{ 2 \alpha_{h} \lambda}{3}}\left[{}_{1}F_{1}\left(1;\frac{3}{2};\frac{\alpha_{h}\lambda}{2}\right) 
    +{}_{1}F_{1}\left(1;\frac{3}{2};\frac{\alpha_{h}\lambda}{6}\right)\right] \nonumber \\
    &+ \frac{2 B_c \alpha_h}{9\pi}e^{-\frac{ 2 \alpha_{h} \lambda}{3}}{}_{1}F_{1}\left(1;\frac{3}{2};\frac{\alpha_{h}\lambda}{3}\right),
\end{align}
where the modulation order dependent terms $\alpha_h, B,$ and $B_c$ can be expressed as $\alpha_h = \frac{24}{7M-4}, B_c = 6\left[1-\frac{1}{\sqrt{M}}\right]^2, B = 2\left[3-\frac{4}{\sqrt{M}}+\frac{1}{M}\right]$. 
Furthermore, by substituting eqs. (\ref{HQAM_Cond_ASER}) and (\ref{CDF_Expression}) into eq. (\ref{ASER_Definition}), the \ac{aser} expression for the \ac{hqam} scheme can be expressed in terms of $\mathcal{I}_1\left(\cdot, \cdot\right)$, $ \mathcal{I}_{2}\left(\cdot, \cdot\right)$, $\mathcal{I}_3\left(\cdot, \cdot, \cdot\right)$, and $\mathcal{I}_{4}\left(\cdot, \cdot, \cdot\right)$ as expressed in eq. \eqref{HQAM_Intermediate_1}.
\begin{figure*}[!t]
\begin{align}\label{HQAM_Intermediate_1}
    P_s^H\left(e\right) = 
   &-\sqrt{\frac{\alpha_h}{2\pi}}\left(\frac{B_c-B}{2}\right)\mathcal{I}_{1}\left(-\frac{1}{2}, \frac{\alpha_h}{2}\right)
    + \sqrt{\frac{\alpha_h}{3\pi}}\left(\frac{B_c}{3}\right)\mathcal{I}_{1}\left(-\frac{1}{2}, \frac{\alpha_h}{3}\right)
    - \sqrt{\frac{\alpha_h}{6\pi}}\left(\frac{B_c}{2}\right)\mathcal{I}_{1}\left(-\frac{1}{2}, \frac{\alpha_h}{6}\right)
    - \frac{2B_c\alpha_h}{9\pi}\mathcal{I}_{3}\left(0,\frac{2\alpha_h}{3}, \frac{\alpha_h}{3}\right) \nonumber \\
   &+ \frac{B_c\alpha_h}{2\sqrt{3}\pi}\!\left[\mathcal{I}_{3}\!\left(0,\frac{2\alpha_h}{3}, \frac{\alpha_h}{2}\right)
    + \mathcal{I}_{3}\!\left(0,\frac{2\alpha_h}{3}, \frac{\alpha_h}{6}\right)\right]
    +\frac{1}{\Gamma\!\left(\tau+1\right)}\!\Bigg[\!\sqrt{\frac{\alpha_h}{2\pi}}\!\left(\frac{B_c-B}{2}\right)\!\mathcal{I}_2\!\left(-\frac{1}{2}, \frac{\alpha_h}{2}\right)
    - \sqrt{\frac{\alpha_h}{3\pi}}\!\left(\frac{B_c}{3}\right)\!\mathcal{I}_2\!\left(-\frac{1}{2}, \frac{\alpha_h}{3}\right) \nonumber \\
   &+ \sqrt{\frac{\alpha_h}{6\pi}}\!\left(\frac{B_c}{2}\right)\!\mathcal{I}_2\!\left(-\frac{1}{2},\frac{\alpha_h}{6}\right)
    + \frac{2B_c\alpha_h}{9\pi}\!\mathcal{I}_4\!\left(0,\frac{2\alpha_h}{3}, \frac{\alpha_h}{3}\right)
    - \frac{B_c\alpha_h}{2\sqrt{3}\pi}\!\mathcal{I}_4\!\left(0, \frac{2\alpha_h}{3}, \frac{\alpha_h}{2}\right)
    - \frac{B_c\alpha_h}{2\sqrt{3}\pi}\!\mathcal{I}_4\!\left(0,\frac{2\alpha_h}{3}, \frac{\alpha_h}{6}\right)\!\Bigg].
\end{align}
\rule{\textwidth}{0.4pt}
\end{figure*}
Further, by substituting the solutions of $\mathcal{I}_{1}\left(\cdot, \cdot\right)$, $\mathcal{I}_2\left(\cdot, \cdot\right)$, $\mathcal{I}_{3}\left(\cdot, \cdot, \cdot\right)$ and $\mathcal{I}_{4}\left(\cdot, \cdot, \cdot\right)$ from Appendices \ref{Appendix_I1}, \ref{Appendix_I2}, \ref{Appendix_I3}, and \ref{Appendix_I4} into eq. (\ref{HQAM_Intermediate_1}), the generalized \ac{aser} expression for the \ac{hqam} scheme can be obtained as expressed in eq. \eqref{HQAM_ASER_Expression}.
\begin{figure*}[t]
\begin{align}\label{HQAM_ASER_Expression}
    P_s^H\left(e\right) =& \frac{B}{2}-\frac{B_c}{3}
    + \frac{1}{\Gamma\left(\tau+1\right)\sqrt{\pi}} \left\{\left(\frac{B_c-B}{2}\right)\right.
    H_{2,2}^{2,1} \left[\frac{\sqrt{2}}{\Lambda\sqrt{\alpha_h\lambda_0}} \bigg| 
    \begin{array}{cc}
         \nu_0\left(-\frac{1}{2}\right), \nu_1 \\
         \nu_2, \nu_3
    \end{array}\right]-\left(\frac{B_c}{3}\right)
    H_{2,2}^{2,1} \left[\frac{\sqrt{3}}{\Lambda\sqrt{\alpha_h\lambda_0}} \bigg| 
    \begin{array}{cc}
         \nu_0\left(-\frac{1}{2}\right), \nu_1 \\
         \nu_2, \nu_3
    \end{array}\right] \nonumber \\
    &+ \left(\frac{B_c}{2}\right) \left. H_{2,2}^{2,1} \left[\frac{\sqrt{6}}{\Lambda\sqrt{\alpha_h\lambda_0}} \bigg| 
    \begin{array}{cc}
         \nu_0\left(-\frac{1}{2}\right), \nu_1 \\
         \nu_2, \nu_3
    \end{array}\right] \right\} + \frac{B_c}{\pi\Gamma\left(\tau+1\right)}
    \left\{\frac{1}{3}H_{1,0:1,2;1,2}^{0,1:1,1;2,0}\left[\begin{array}{cc}
        1  \\
        \frac{\sqrt{3}}{\Lambda\sqrt{\alpha_h\lambda_0}}
             \end{array}\left|\begin{array}{c}
                \nu_6\left(0\right):\nu_7;\nu_1\hfill  \\
                -:\nu_3, \nu_8;\nu_2,\nu_3 
             \end{array}\right.\right] \right. \nonumber \\
& - \frac{\sqrt{3}}{2}H_{1,0:1,2;1,2}^{0,1:1,1;2,0}\left[\begin{array}{cc}
         3  \\
        \frac{\sqrt{6}}{\Lambda\sqrt{\alpha_h\lambda_0}}
             \end{array}\left|\begin{array}{c}
                \nu_6\left(0\right):\nu_7;\nu_1\hfill  \\
                -:\nu_3, \nu_8;\nu_2,\nu_3 
             \end{array}\right.\right] 
 - \left.\frac{1}{2\sqrt{3}}H_{1,0:1,2;1,2}^{0,1:1,1;2,0}\left[\begin{array}{cc}
        1/3  \\
        \frac{\sqrt{2}}{\Lambda\sqrt{\alpha_h\lambda_0}}
             \end{array}\left|\begin{array}{c}
                \nu_6\left(0\right):\nu_7;\nu_1\hfill  \\
                -:\nu_3, \nu_8;\nu_2,\nu_3 
             \end{array}\right.\right] \right\}             
\end{align}
\rule{\textwidth}{0.4pt}
\end{figure*}

\textbf{Remark 2:} The closed-form \ac{aser} expressions for both \ac{rqam} and \ac{hqam} schemes, given in \eqref{rqam_aser_intermediate_2} and \eqref{HQAM_ASER_Expression}, are represented in terms of univariate Fox–H function $H_{2,2}^{2,1} \!\left[\frac{\chi_1}{\Lambda \sqrt{\lambda_0}} \Bigg| 
\begin{array}{cc} \nu_0\!\left(-0.5\right), \nu_1 \\ (\tau+1,1), \nu_3 \end{array}\!\right]$ and the bivariate Fox–H function $H_{1,0:1,2;1,2}^{0,1:1,1;2,0}\!\left[\begin{array}{cc} \chi_2 \\ \tfrac{\chi_3}{\Lambda\sqrt{\lambda_0}} \end{array}\!\Bigg| \begin{array}{c} \nu_6(0):\nu_7;\nu_1 \\ -:\nu_3,\nu_8;(\tau+1,1),\nu_3 \end{array}\!\right]$, where $\chi_1, \chi_2, \chi_3$ are modulation order specific constants. It is observed that the values of both the univariate and bivariate Fox–H functions increase monotonically with respect to $\Lambda$, $\tau$, and $\lambda_0$, indicating that these functions exhibit positive monotonicity with respect to the aforementioned variables. Furthermore, the bivariate Fox–H function dominates over its univariate counterpart. Consequently, as $\Lambda$, $\tau$, and $\lambda_0$ increase, the overall \ac{aser} values decrease. Therefore, it can be inferred that an increase in the fading parameters, pointing error parameters, and $N$ results in an improvement in the \ac{aser} performance for both the \ac{rqam} and \ac{hqam} schemes.
%
\subsection{Channel Capacity}
The \ac{acc} is defined as the maximum achievable rate of reliable data transmission over a fading channel. For any system model, it can be derived by using \ac{cdf}-based approach \cite[eq. (27)]{9079579} as
\begin{align}\label{cc_Def}
    \bar{\eta} = \frac{1}{ln_{e}\left(2\right)}\int_{\lambda=0}^\infty \frac{1-F_{\lambda}\left(\lambda\right)}{1+\lambda} d\lambda. 
\end{align}
Further, the \ac{cdf} in eq. (\ref{CDF_Expression}) is reformulated using the Meijer-G function \cite[eq. (8.356.3)]{gradshteyn2014table}. By substituting this expression into eq. (\ref{cc_Def}), $\bar{\eta}$ can be obtained in terms of integral function $\mathcal{I}_{5}$ as
\begin{align}\label{CC_Intermediate_2}
    \bar{\eta} = \frac{1}{ln_{e}\left(2\right)}\frac{1}{\Gamma\left(\tau+1\right)}\mathcal{I}_{5},
\end{align}
where $\mathcal{I}_{5}$ can be expressed as
\begin{align}\label{I5_Def}
    \mathcal{I}_{5} = \int_{\lambda=0}^\infty G_{1,1}^{1,1}\left[\lambda\left|\begin{array}{cc}
         0  \\
         0 
    \end{array}\right.\right]\Gamma\left(\tau+1, \frac{\sqrt{\lambda}}{\Lambda\sqrt{\lambda_{0}}}\right) d\lambda.
\end{align}
The integral $\mathcal{I}_{5}$ is evaluated in appendix \ref{Appendix_I5}. Accordingly, based on eq. (\ref{I5_Final}), $\bar{\eta}$ can be represented in terms of the univariate Fox‑H function as
\begin{align}\label{CC_Final}
    \bar{\eta} = \frac{1}{ln_{e}\left(2\right)} \frac{1}{\Gamma\left(\tau+1\right)}H_{3,2}^{1,3}\left[\Lambda^2\lambda_0\left|\begin{array}{cc}
         \nu_1, \nu_1, \nu_9  \\
         \nu_1, \nu_3
    \end{array}\right.\right].
\end{align}
\textbf{Remark 3:}
The closed-form expression of the \ac{acc} provided in eq.~(\ref{CC_Final}), represented in terms of the Fox–H function
$H_{3,2}^{1,3}\left[\Lambda^2\lambda_0\left|\begin{array}{cc} \nu_1, \nu_1, \left(-\tau,2\right) \\ \nu_1, \nu_{3} \end{array}\right.\right]$,
characterizes the behavior of $\bar{\eta}$. An increase in either $\Lambda^2\lambda_0$ or $\tau$ results in a corresponding enhancement in the Fox–H function value. Mathematically, the parameter $\tau$ is directly proportional to $\alpha$, $\mu$, $\phi$, $S_0$, and $N$, while remaining independent of $S_0$. Conversely, $\Lambda^2$ varies inversely with $\alpha$ and $\mu$, exhibits direct proportionality with $\phi$ and $S_0$, and remains independent of $N$. However, the influence of $\tau$ is dominant over that of $\Lambda^2$. Therefore, the observed improvement in the \ac{acc} with increasing $\alpha$, $\mu$, $\phi$, $S_0$, $N$, and $\lambda_0$ can be analytically justified through the behavior of the Fox–H function incorporated in eq.~(\ref{CC_Final}).

\subsection{Asymptotic Analysis}
To investigate the system performance in the high \ac{snr} regime, we analyze the asymptotic behavior of the \ac{e2e} \ac{cdf} of the instantaneous \ac{snr} $\lambda$. At high \ac{snr}, the \ac{cdf} of $\lambda$ is approximated as
\begin{align} \label{Asymptotic e2e snr 1}
    F_{\lambda}^{\infty}(\lambda) = \frac{1}{\Gamma(\tau+1)}\lim_{\lambda_0 \to\infty} \gamma\left(\tau+1, \frac{\sqrt{\lambda}}{\Lambda\sqrt{\lambda_0}}\right),
\end{align}
Utilizing its asymptotic series expansion \cite[eq. (06.06.06.0002.01)]{wolfram_functions} and retaining only the first dominant term, we obtain the following approximation:
\begin{align} \label{Asymptotic e2e snr 2}
    F_{\lambda}^{\infty}(\lambda) \approx \frac{1}{\Gamma(\tau+2)}\left(\frac{\sqrt{\lambda}}{\Lambda\sqrt{\lambda_0}}\right)^{\tau+1}.
\end{align}

\subsubsection{Asymptotic Outage Probability}
Consequently, the asymptotic \ac{op} can be directly expressed by using eq. (\ref{Asymptotic e2e snr 2}) as
\begin{align} \label{Asymptotic_Outage_Probability}
    P_O^\infty(\lambda_{th}) = F_{\lambda}^{\infty}(\lambda_{th}) \approx \frac{1}{\Gamma(\tau+1)}\left(\frac{\lambda_{th}}{\Lambda^{2}\lambda_0}\right)^{\frac{\tau+1}{2}}.
\end{align}  
Further, eq. (\ref{Asymptotic_Outage_Probability}) can also be represented in terms of diversity order $\left(G_d\right)$ and the coding gain $(G_c^{\mathrm{op}})$ as \cite[eq. (21)]{HINDUSTANI_IRS}
\begin{align}\label{DO_CD_Def_From_OP}
     P_O^\infty(\lambda_{th}) \approx \frac{1}{\left(G_c^{\mathrm{op}} \bar{\lambda}_0\right)^{G_d}},
\end{align}
where, the $G_d$ and $G_c^{op}$ can be expressed as
\begin{align}
    G_d      &= \frac{\tau+1}{2} \label{gd_from_Op} \\
    G_c^{op} &= \left[\frac{1}{\Gamma\left(\tau+1\right)}\frac{\lambda_{th}}{\Lambda^2}\right]^{-\frac{1}{G_d}}.
\end{align}
Given that $\tau = \frac{\mu_{\mathcal{B}}^2}{\sigma_{\mathcal{B}}^2} - 1$, where $\mu_{\mathcal{B}} = \sum_{j=1}^N \mu_{h_j}$ and $\sigma_{\mathcal{B}}^2 = \sum_{j=1}^N \sigma_{h_j}^2$ are defined in Section \ref{sub_cdf}, and the expressions for $\mu_{h_j}$ and $\sigma^2_{h_j}$ are provided in eqs. \eqref{IRS_Mean} and \eqref{IRS_Variance}, respectively. By substituting these into eq. \eqref{gd_from_Op}, the modified form of the diversity gain is obtained as
\begin{align}\label{Simplified_Gd_Expression}
    G_d = \frac{N}{2}\left\{\prod_{i=1}^{2}\frac{\Gamma\left(\frac{2}{\alpha_{ij}}+\mu_{ij}\right)\Gamma\left(\mu_{ij}\right)\left(\phi_{ij}+1\right)^{2}}{\left\{\Gamma\left(\frac{1}{\alpha_{ij}}+\mu_{ij}\right)\right\}^{2}\phi_{ij}\left(\phi_{ij}+2\right)}-1\right\}^{-1}.
\end{align}
\emph{\textbf{Remark 4:} It can be observed that $G_d$ depends only on $N$, $\alpha_{ij}$, $\mu_{ij}$, and $\phi_{ij}$.}

\subsubsection{Asymptotic Average Symbol Error Rate}
By substituting eqs. \eqref{Asymptotic e2e snr 2} and \eqref{RQAM_Cond_ASER} into eq. \eqref{ASER_Definition}, asymptotic \ac{aser} for \ac{rqam} scheme can be expressed in terms of $\mathcal{I}_{1}\left(\cdot, \cdot\right)$ and $\mathcal{I}_{3}\left(\cdot, \cdot, \cdot\right)$ as
\begin{align}\label{Asymptotic_ASER}
    P_s^\infty\left(e\right) & \approx \frac{1}{\Gamma\left(\tau+2\right)}\left(\frac{1}{\Lambda\sqrt{\lambda_0}}\right)^{\tau+1}\left[-\mathcal{DI}_1\left(\frac{\tau}{2}, \frac{a^2}{2}\right) \right.\nonumber \\
    & - \mathcal{FI}_1\left(\frac{\tau}{2}, \frac{b^2}{2}\right) + \frac{\mathcal{G}}{\sqrt{\pi}}\left(\mathcal{I}_3\left(\frac{\tau+1}{2}, \frac{a^2+b^2}{2}, \frac{a^2}{2}\right) \right.\nonumber \\
    & \left.\left.+ \mathcal{I}_3\left(\frac{\tau+1}{2}, \frac{a^2+b^2}{2}, \frac{b^2}{2}\right)\right)\right].
\end{align}
Furthermore, by substituting the solutions of $\mathcal{I}_1\left(\cdot, \cdot\right)$ and $\mathcal{I}_3\left(\cdot, \cdot, \cdot\right)$ from appendices \ref{Appendix_I1} and \ref{Appendix_I3}, respectively, into eq. (\ref{Asymptotic_ASER}), the asymptotic \ac{aser} can be obtained as
\begin{align}\label{Asymptotic ASER Final}
    P_s^\infty\left(e\right) & \approx \frac{1}{\Gamma\left(\tau+2\right)}\left(\frac{1}{\Lambda\sqrt{\lambda_0}}\right)^{\tau+1}\left[-\mathcal{D}\frac{\Gamma\left(\frac{\tau}{2}+1\right)}{\left(\frac{a^2}{2}\right)^{\frac{\tau}{2}+1}} +\frac{\mathcal{G}}{\sqrt{\pi}}\times\right.\nonumber \\
    & \frac{\Gamma\left(\frac{\tau+3}{2}\right)}{\left(\frac{a^2+b^2}{2}\right)^{\frac{\tau+3}{2}}} \times\left({}_{2}F_{1}\left(1,\frac{\tau+1}{2}+1;\frac{3}{2};\frac{a^2}{a^2+b^2}\right)\right. \nonumber \\
    & \left.+{}_{2}F_{1}\left(1,\frac{\tau+1}{2}+1;\frac{3}{2};\frac{b^2}{a^2+b^2}\right)\right) \left. - \mathcal{F}\frac{\Gamma\left(\frac{\tau}{2}+1\right)}{\left(\frac{b^2}{2}\right)^{\frac{\tau}{2}+1}}\right].
\end{align}
With the aid of \cite[eq. (17)]{dixit2017exact}, the asymptotic \ac{aser} in eq. (\ref{Asymptotic ASER Final}) can be expressed as
\begin{align}\label{Asymptotic_Form}
    P_s^\infty\left(e\right) \approx \frac{1}{\left(G_c^R\bar{\lambda}_0\right)^{G_d}}. 
\end{align}
Here, $G_d$ and $G_c^R$ represent the diversity and coding gains for \ac{rqam} scheme, respectively, and are expressed as follows:
\begin{align}
    G_d &= \frac{\tau+1}{2}, \label{Gd_Expression_from_aser} \\
    G_c^R &= \left[\frac{\Lambda^{-\tau-1}}{\Gamma\left(\tau+2\right)}\left\{-\mathcal{D}\frac{\Gamma\left(\frac{\tau}{2}+1\right)}{\left(\frac{a^2}{2}\right)^{\frac{\tau}{2}+1}} - \mathcal{F}\frac{\Gamma\left(\frac{\tau}{2}+1\right)}{\left(\frac{b^2}{2}\right)^{\frac{\tau}{2}+1}} + \frac{\mathcal{G}}{\sqrt{\pi}} \right.\right. \nonumber \\
    & \times \frac{\Gamma\left(\frac{\tau+3}{2}\right)}{\left(\frac{a^2+b^2}{2}\right)^{\frac{\tau+3}{2}}} \left({}_{2}F_{1}\left(1,\frac{\tau+1}{2}+1;\frac{3}{2};\frac{a^2}{a^2+b^2}\right)\right. \nonumber \\
    & \left.\left.\left.+{}_{2}F_{1}\left(1,\frac{\tau+1}{2}+1;\frac{3}{2};\frac{b^2}{a^2+b^2}\right)\right)\right\}\right]^{-\frac{1}{G_d}}. \label{Gc_Expression}
\end{align}
It can be observed that the expressions of $G_d$ obtained from eqs. \eqref{gd_from_Op} and \eqref{Gd_Expression_from_aser} are the same, as expected.

\section{DNN-based Performance Analysis}\label{DNN Section}
Analytical evaluation of performance metrics often involves generalized functions such as the incomplete Gamma, Meijer~G, and Fox–H, whose numerical computation is intensive for real-time use on resource-constrained systems. Hence, equivalent \ac{dnn}-based frameworks are proposed to accurately predict the \ac{op} and \ac{aser} for the considered \ac{rqam} scheme, as detailed in the following subsections.

\subsection{Design of the DNN Model to Predict OP}\label{DNN_OP}
\begin{figure}[t!]
    \centering
    \includegraphics[width=1\linewidth]{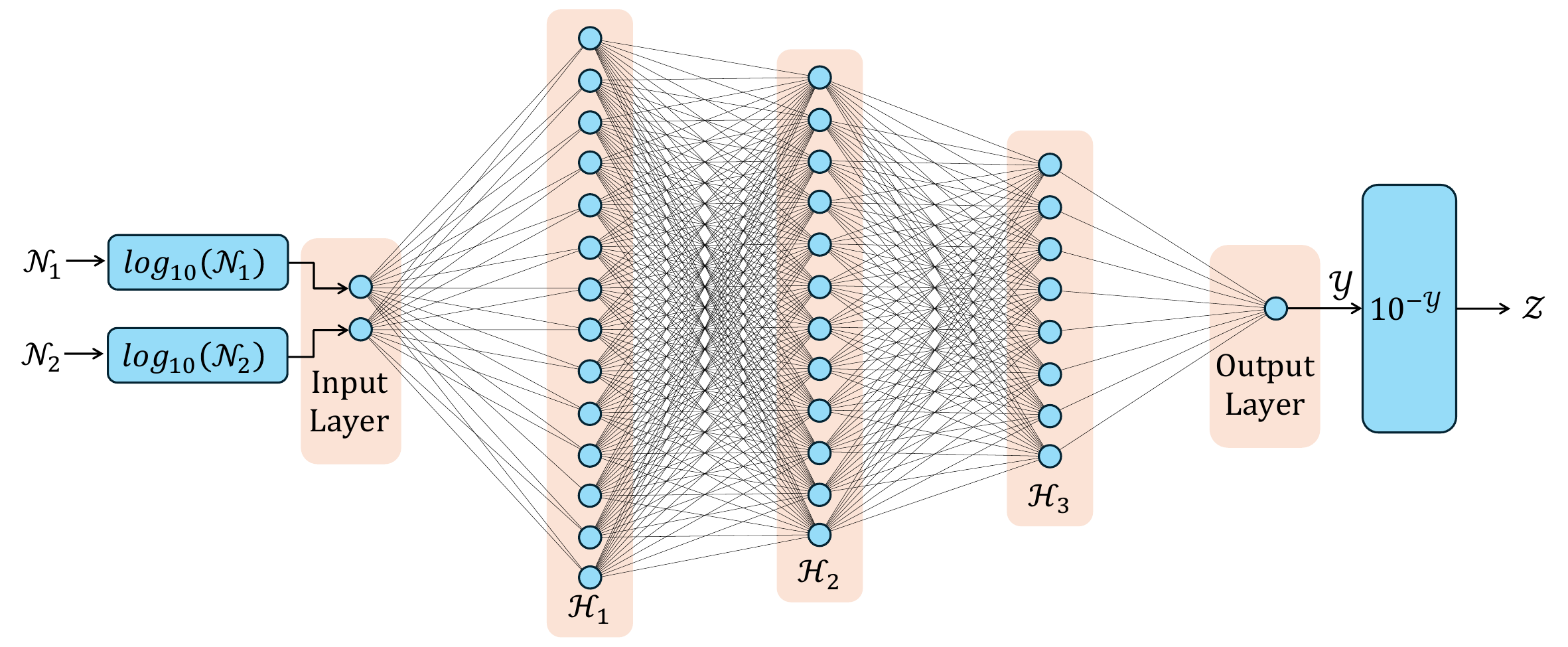}
    \caption{The \ac{dnn} architecture for \ac{op} prediction of the proposed system model}
    \label{fig:DNN_OP}
\end{figure}

In this subsection, a \ac{dnn} model is developed to predict the \ac{op} of the proposed \ac{irs}-assisted wireless communication system. The network is implemented as a \ac{ffn} using the MATLAB Neural Network Toolbox. The architecture comprises an input layer with two neurons, three fully connected hidden layers, $\mathcal{H}_1$, $\mathcal{H}_2$, and $\mathcal{H}_3$, containing 14, 12, and 8 neurons, respectively, and a single-neuron output layer $(\mathcal{Y})$, as illustrated in Fig.~\ref{fig:DNN_OP}. 

\begin{figure}[t!]
    \centering
    \includegraphics[width=9cm, height=6cm]{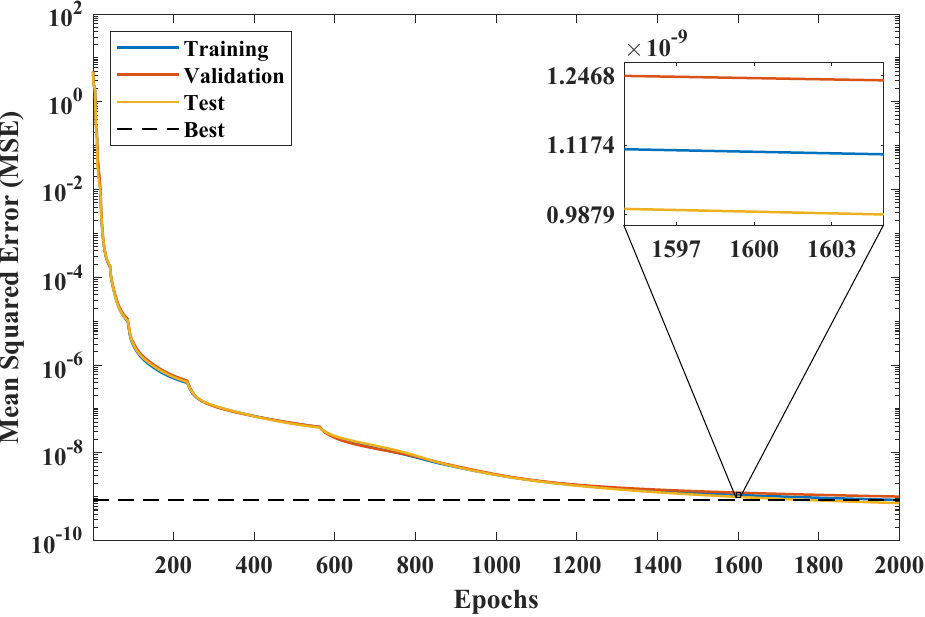}
    \caption{Training process analysis for proposed \ac{dnn}-based \ac{op} prediction method}
    \label{fig:PlotPerform_OP}
\end{figure}

\begin{figure}[t!]
    \centering
    \includegraphics[width=9cm, height=6cm]{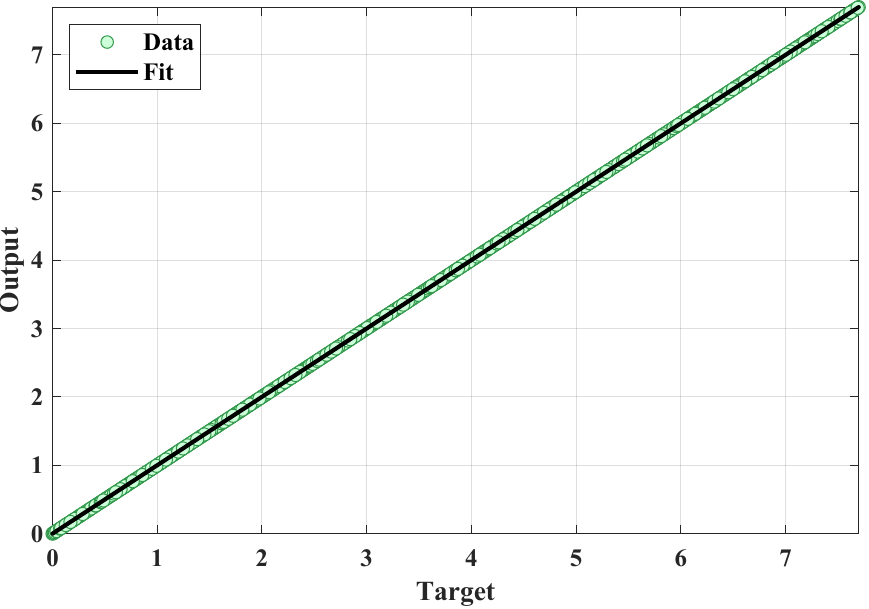}
    \caption{Regression analysis result for \ac{op} prediction}
    \label{fig:PlotRegression_OP}
\end{figure}

The hidden layers employ the default nonlinear transfer functions of MATLAB’s \texttt{feedforwardnet} implementation. To improve numerical conditioning, both input features $(\mathcal{N}_1, \mathcal{N}_2)$ are transformed using the base-10 logarithm, while the predicted output is obtained by applying $10^{-(\cdot)}$ to the network output. The inputs and the predicted output of \ac{op}, $P_O^{\mathrm{pred}}(\lambda_{th})$, can be expressed as
\begin{align}\label{DNN_Input_OP}
    \mathcal{N}_1 &= \tau+1 \\
    \mathcal{N}_2 &= \frac{\lambda_{th}}{\mu_\mathcal{B}\sqrt{\lambda_0}} \\
    P_O^{\mathrm{pred}}\left(\lambda_{th}\right) &= \mathcal{Z} = 10^{-\mathcal{Y}}.
\end{align}

The training dataset is generated by evaluating $4\times 10^5$ distinct parameter combinations of $\tau$, $\lambda_{th}$, $\mu_\mathcal{B}$, and $\lambda_0$ using the expression provided in eq. \eqref{Outage_Probability}. The MATLAB training routine randomly assigns $70\%$ of the dataset to training, while the remaining $30\%$ is equally divided into validation and testing subsets. The network is then subsequently trained in a supervised manner using the Levenberg--Marquardt backpropagation algorithm (\texttt{trainlm}) for up to $2000$ epochs. To mitigate overfitting, early stopping based on validation performance is employed.

Fig.~\ref{fig:PlotPerform_OP} depicts the training progression of the \ac{dnn}-based \ac{op} prediction model, including the \ac{mse} trajectories for all phases, the total number of epochs, and the overall training time. As no increase in validation error is observed, the training continues to the maximum limit of $2000$ epochs, while a minimum validation \ac{mse} of $8.5\times 10^{-10}$ is observed. The relationship between the actual \ac{op} results obtained using eq.~\eqref{Outage_Probability} and the corresponding \ac{dnn} predictions is presented in Fig.~\ref{fig:PlotRegression_OP}. The test-phase intercept of $8.6 \times 10^{-7}$ indicates a negligible offset between predicted and true values. Furthermore, the correlation coefficient, approximately $1$, indicates an almost perfect agreement between the true and predicted results. Furthermore, the comparison of the calculation time and the \ac{mse} between the actual results and the \ac{dnn}-predicted results is summarized in Table~\ref{tab:OP MSE time table}.

\begin{table}[t]
\centering
\caption{Execution time and \ac{mse} in the case of $1000$ samples using analytical \ac{op} expression.}
\label{tab:OP MSE time table}
\begin{tabular}{|l|c|c|}
\hline
\textbf{Parameter} & \textbf{Exact expression} & \textbf{\ac{dnn}-based prediction} \\
\hline
MSE     & 0        & $8.5\times10^{-10}$  \\
\hline
Time (s) & 1832.6   & 1.284              \\
\hline
\end{tabular}
\end{table}

\subsection{DNN-Based Framework for ASER Estimation}\label{DNN_ASER}
\begin{figure}[t!]
    \centering
    \includegraphics[width=1\linewidth]{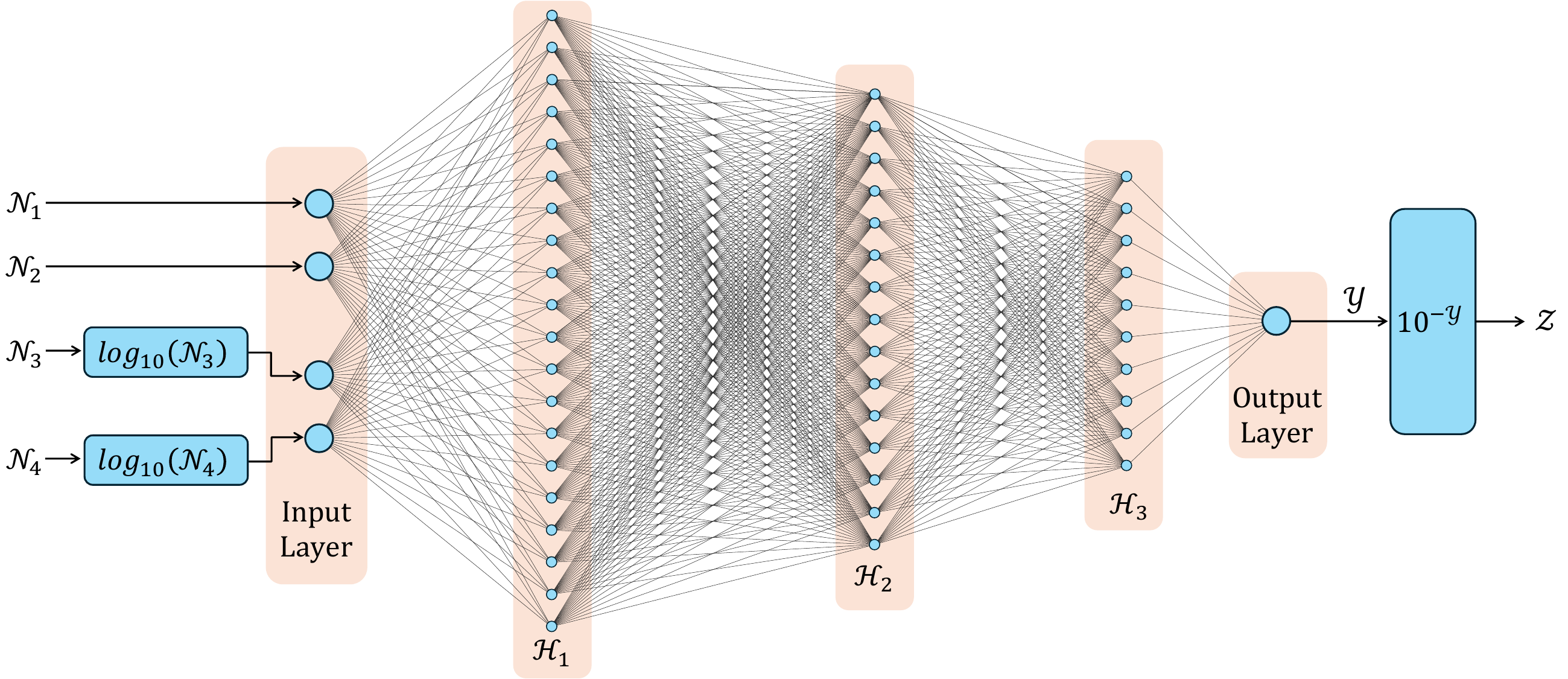}
    \caption{The \ac{dnn} architecture for \ac{aser} prediction of the \ac{rqam} scheme}
    \label{fig:DNN_ASER}
\end{figure}

To estimate the \ac{aser} for \ac{rqam} scheme of the proposed \ac{irs}-assisted wireless communication setup, a \ac{dnn} model is designed and implemented as a \ac{ffn} using MATLAB’s Neural Network Toolbox. The architecture, illustrated in Fig.~\ref{fig:DNN_ASER}, consists of an input stage with four neurons representing $\mathcal{N}_1$, $\mathcal{N}_2$, $\mathcal{N}_3$, and $\mathcal{N}_4$, followed by three fully connected hidden layers $\mathcal{H}_1$, $\mathcal{H}_2$, and $\mathcal{H}_3$ containing 20, 15, and 10 neurons, respectively, and a single output neuron $\mathcal{Y}$. All hidden layers employ the default nonlinear activation functions available in the \texttt{feedforwardnet} structure. 

\begin{figure}[t!]
    \centering
    \includegraphics[width=9cm, height=6cm]{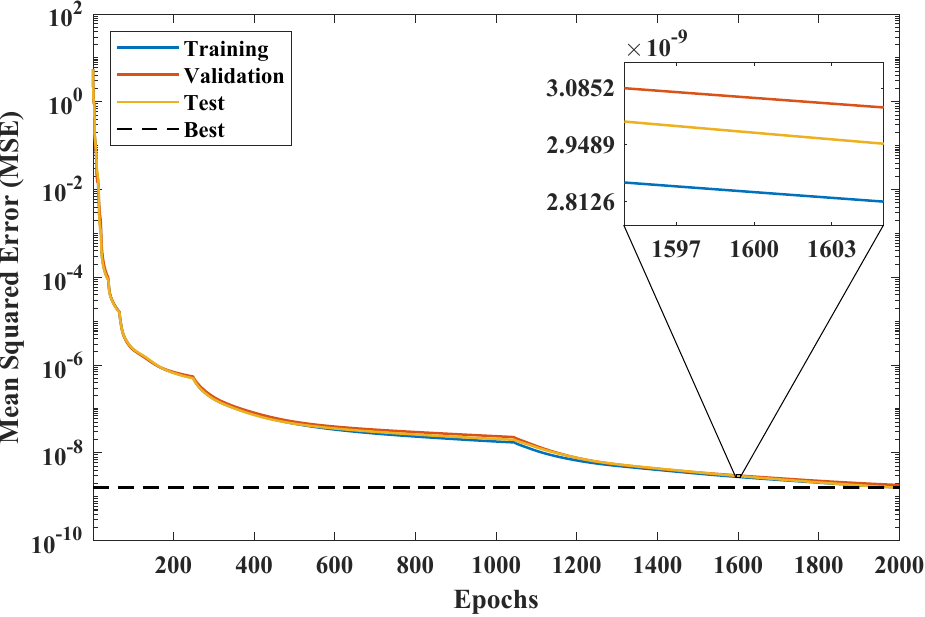}
    \caption{Training process analysis for proposed \ac{dnn}-based \ac{aser} prediction method for \ac{rqam} scheme}
    \label{fig:PlotPerform_RQAM}
\end{figure}

\begin{figure}[t!]
    \centering
    \includegraphics[width=9cm, height=6cm]{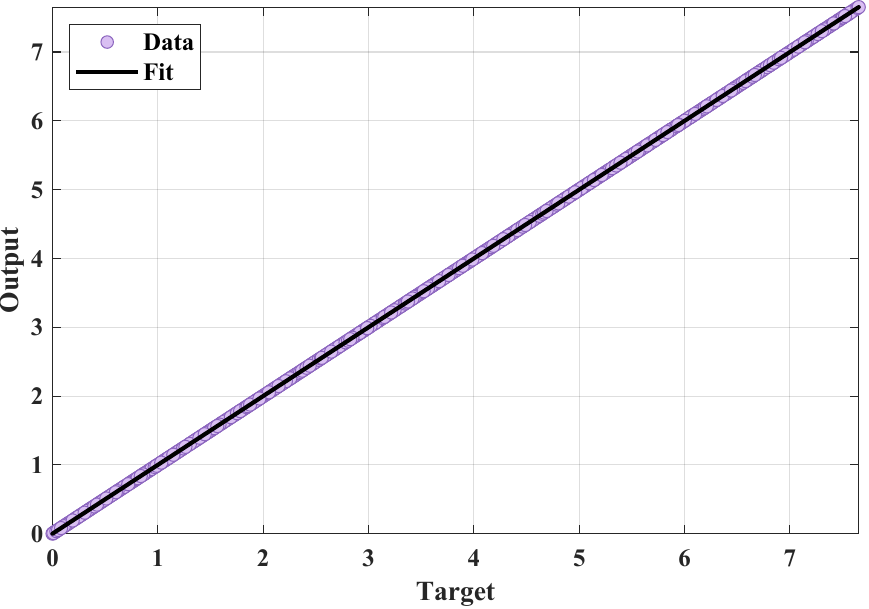}
    \caption{Regression analysis results of \ac{aser} for \ac{rqam} prediction}
    \label{fig:PlotRegression_RQAM}
\end{figure}

The model inputs are defined as $\mathcal{N}_1 = M_I$, $\mathcal{N}_2 = M_Q$, $\log_{10}(\mathcal{N}_3) = \log_{10}(\tau+1)$, and $\log_{10}(\mathcal{N}_4) = \log_{10}(\Lambda\sqrt{\lambda_0})$, where the logarithmic transformation of $\tau+1$ and $\Lambda\sqrt{\lambda_0}$ are introduced to improve numerical conditioning. The network output $\mathcal{Y}$ is post-processed using $\mathcal{Z} = 10^{-\mathcal{Y}}$ to yield the estimated ASER. The relationships are summarized as:
\begin{align}\label{DNN_Input_RQAM}
\mathcal{N}_1 &= M_I \\
\mathcal{N}_2 &= M_Q \\
\mathcal{N}_3 &= \tau+1 \\
\mathcal{N}_4 &= \Lambda\sqrt{\lambda_0} \\
\mathrm{ASER}^{\mathrm{pred}} &= \mathcal{Z} = 10^{-\mathcal{Y}}.
\end{align}

The dataset for training and evaluation is prepared by computing $102400$ different parameter sets for $M_I$, $M_Q$, $\tau+1$, and $\Lambda\sqrt{\lambda_0}$. The network is trained in a supervised manner using the Levenberg--Marquardt backpropagation algorithm (\texttt{trainlm}), with a maximum of $2000$ epochs. For training, we consider $102400$ distinct combinations in the dataset. Wherein all combinations of $M_I$ and $M_Q$ individually take values from $\{2,4,8,16\}$. Both parameters $\tau+1$ and $\Lambda\sqrt{\lambda_0}$ are independently varied within the range between $10^{-7}$ to $200$. During the training in MATLAB, $70\%$ of the dataset is allocated for training, while the remaining $30\%$ is equally divided between validation and testing. To prevent overfitting, MATLAB employs validation-based early stopping, where training halts once the validation error begins to increase, and the network weights corresponding to the minimum validation error are restored.

Fig.~\ref{fig:PlotPerform_RQAM} illustrates the training process of the \ac{dnn}-based \ac{aser} prediction model for the \ac{rqam} scheme. No increase in validation error is observed during training. Hence, the training proceeded until the maximum limit of $2000$ epochs, achieving a minimum validation \ac{mse} of $1.85 \times 10^{-9}$. Furthermore, the relationship between the actual \ac{rqam} \ac{aser} results obtained using eq.~\eqref{rqam_aser_intermediate_2} and the predictions from the \ac{dnn} model is shown in Fig.~\ref{fig:PlotRegression_RQAM}. The test-phase intercept is $7.2 \times 10^{-7}$, indicating a negligible offset between the predicted and actual values. The correlation coefficient is approximately $1$, which is extremely close to perfect correlation. Furthermore, Table~\ref{tab:RQAM ASER MSE time table} reports the comparison of the computation time and the \ac{mse} between the actual \ac{aser} results for \ac{rqam} those predicted by the \ac{dnn} model.

\begin{table}[t]
\centering
\caption{Execution time and \ac{mse} in the case of $1000$ samples using analytical \ac{aser} expression for \ac{rqam} scheme.}
\label{tab:RQAM ASER MSE time table}
\begin{tabular}{|l|c|c|}
\hline
\textbf{Parameter} & \textbf{Exact expression} & \textbf{\ac{dnn}-based prediction} \\ \hline
MSE     & 0        & $1.85\times10^{-9}$  \\
\hline
Time (s) & 9547.2   & 2.276 \\ 
\hline
\end{tabular}
\end{table}

\section{Numerical and simulation results}\label{Numerical and simulation results}
To validate the correctness of the derived expressions, we obtain Monte Carlo simulation results and asymptotic results for the considered \ac{irs}-assisted \ac{thz} band communication. We choose $10^7$ realizations of channel coefficients and symbols in the simulations. The impact of environmental conditions, such as absolute temperature, relative humidity, and atmospheric pressure, is summarized in Table~\ref{Parameter_table}. Furthermore, the ranges of all parameters considered for the Monte Carlo simulations are also listed in Table~\ref{Parameter_table}.
\begin{table}[htb!]
\centering
\caption{List of Simulation Parameters}
\label{Parameter_table}
\resizebox{\columnwidth}{!}{%
\begin{tabular}{|l|l|}
\hline
\textbf{Parameter} & \textbf{Value} \\ \hline
THz Carrier Frequency ($f$) & 275 GHz \\ \hline
\ac{s}-\ac{irs}, \ac{irs}-\ac{d} Link Distance ($d_i$) & 15 m \\ \hline
Transmitter and Receiver Antenna Gain ($G_{t_i}, G_{r_i}$) & 55 dBi \\ \hline
Transmit Power ($P_s$) & -5 to 40 dBm \\ \hline
Noise Variance ($\sigma_n^2$) & 6.08 $\mu$W \\ \hline
Absolute Temperature ($T$) & 296 K \\ \hline
Atmospheric Pressure ($P$) & 1013.25 hPa \\ \hline
Relative Humidity ($\Psi$) & 50\% \\ \hline
THz Link Fading Parameter ($\alpha_i$) & 1 to 3.5 \\ \hline
Normalized Variance of THz Link Envelope ($\mu_i$) & 1 to 3.5 \\ \hline
$\alpha$-root Mean Value of THz Link Envelope ($\Omega_i$) & 1 \\ \hline
Pointing Error Parameter ($\Phi_{i}$) & 4 to 18 \\ \hline
Fraction of Power Collected at Center ($S_{0_{i}}$) & 0.6 to 0.8 \\ \hline
\end{tabular}%
}
\end{table}
In the simulations, the \ac{irs} elements are assumed to be compact and closely spaced. Consequently, all incident links from \ac{s} to the \ac{irs} are modeled with a common distance $d_1$, while all reflected links from the \ac{irs} to \ac{d} are modeled with a common distance $d_2$. Furthermore, the links associated with the $i^{th}$ hop are considered to be \ac{iid}, characterized by fading parameters $\alpha_i$, $\mu_i$, $\Omega_i$, and pointing error coefficients $\phi_i$, and $S_{0_{i}}$.

\begin{figure}[t!]
    \centering
     \includegraphics[width=9cm, height=6cm]{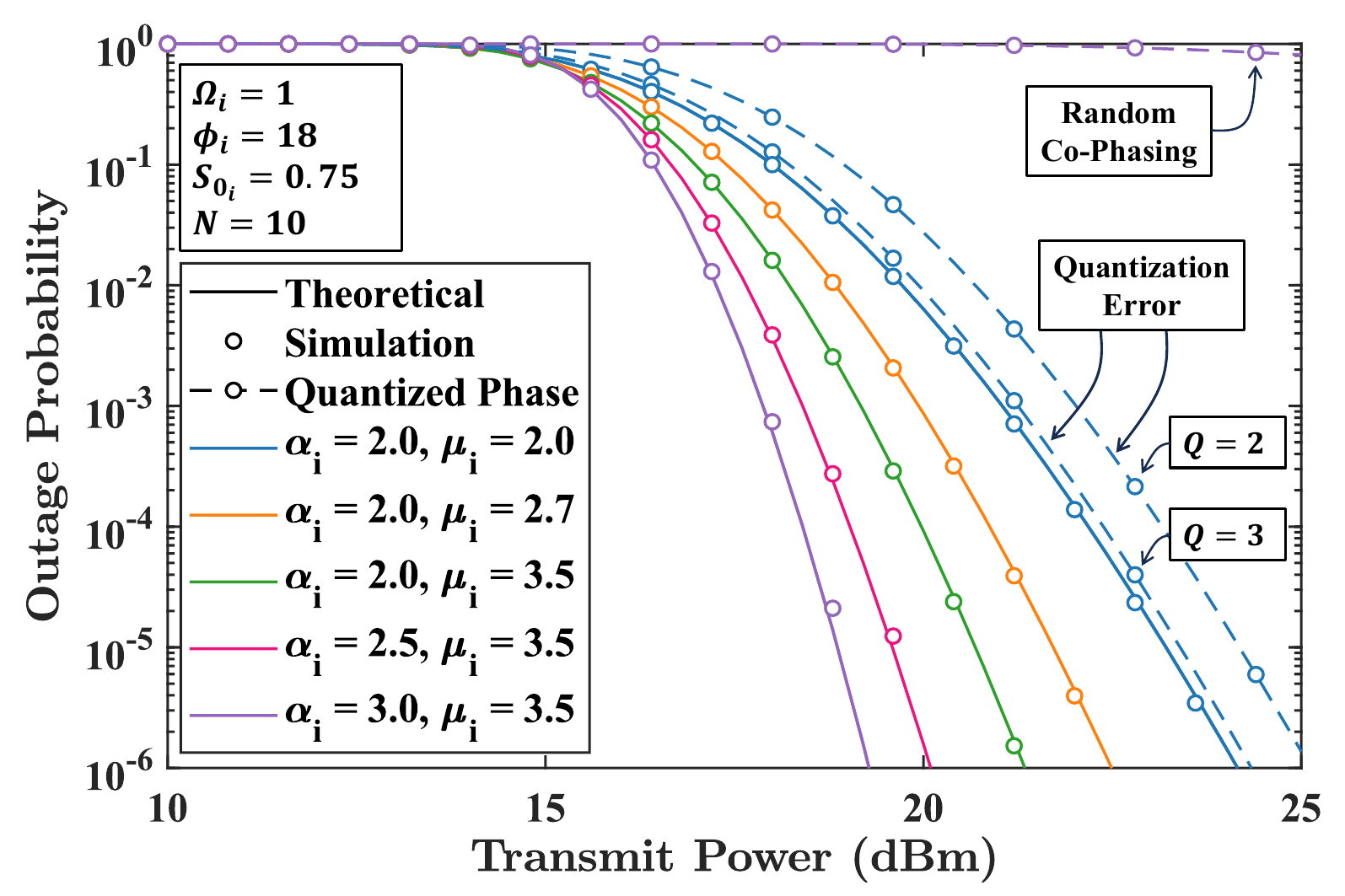}
    \caption{Impact of $\alpha_i$ and $\mu_i$ on the \ac{op} of the \ac{e2e} System Model}
    \label{fig:OP Alpha Mu}
\end{figure}
In Fig. \ref{fig:OP Alpha Mu}, the impact of the fading parameters $\alpha_i$ and $\mu_i$ on the \ac{op} performance is analyzed, while keeping the other parameters fixed as: $\Omega_i = 1$, $\phi_i = 18$, $S_{0_{i}} = 0.75$, and $N = 10$. A noticeable improvement in the \ac{op} is observed with an increase in either $\alpha_i$ or $\mu_i$. For example, when $\mu_i$ is fixed at $3.5$, and $\alpha_i$ is increased from $2$ to $2.5$ and $3$, transmit power reductions of approximately $1.24$ dB and $2.04$ dB, respectively, are required to maintain the \ac{op} at $10^{-6}$. This improvement can be attributed to the fact that larger values of $\alpha_i$ correspond to less severe multipath fading, resulting in more favorable channel conditions. Furthermore, when $\alpha_i$ is fixed at $\alpha_i = 2$, and $\mu_i$ is varied from $2$ to $2.7$ and $3.5$, approximately $1.67$ dB and $2.79$ dB less transmit power is required, respectively, to sustain the same \ac{op} threshold at $10^{-6}$. The observed enhancement is attributed to the increased number of multipath clusters, represented by higher $\mu_i$ values, which contribute to a more stable signal envelope. These results clearly demonstrate the sensitivity of \ac{op} performance to the fading parameters, and highlight the importance of accurately modeling the small-scale fading characteristics in system-level performance evaluation. Furthermore, simulation results incorporating the effects of random cophasing and quantization error are presented. It is observed that random cophasing significantly degrades the \ac{op} performance. In contrast, the degradation caused by quantization error becomes negligible for $Q>3$ when a $Q$-bit quantizer is employed in the \ac{irs}.

\begin{figure}[t!]
    \centering
    \includegraphics[width=9cm, height=6cm]{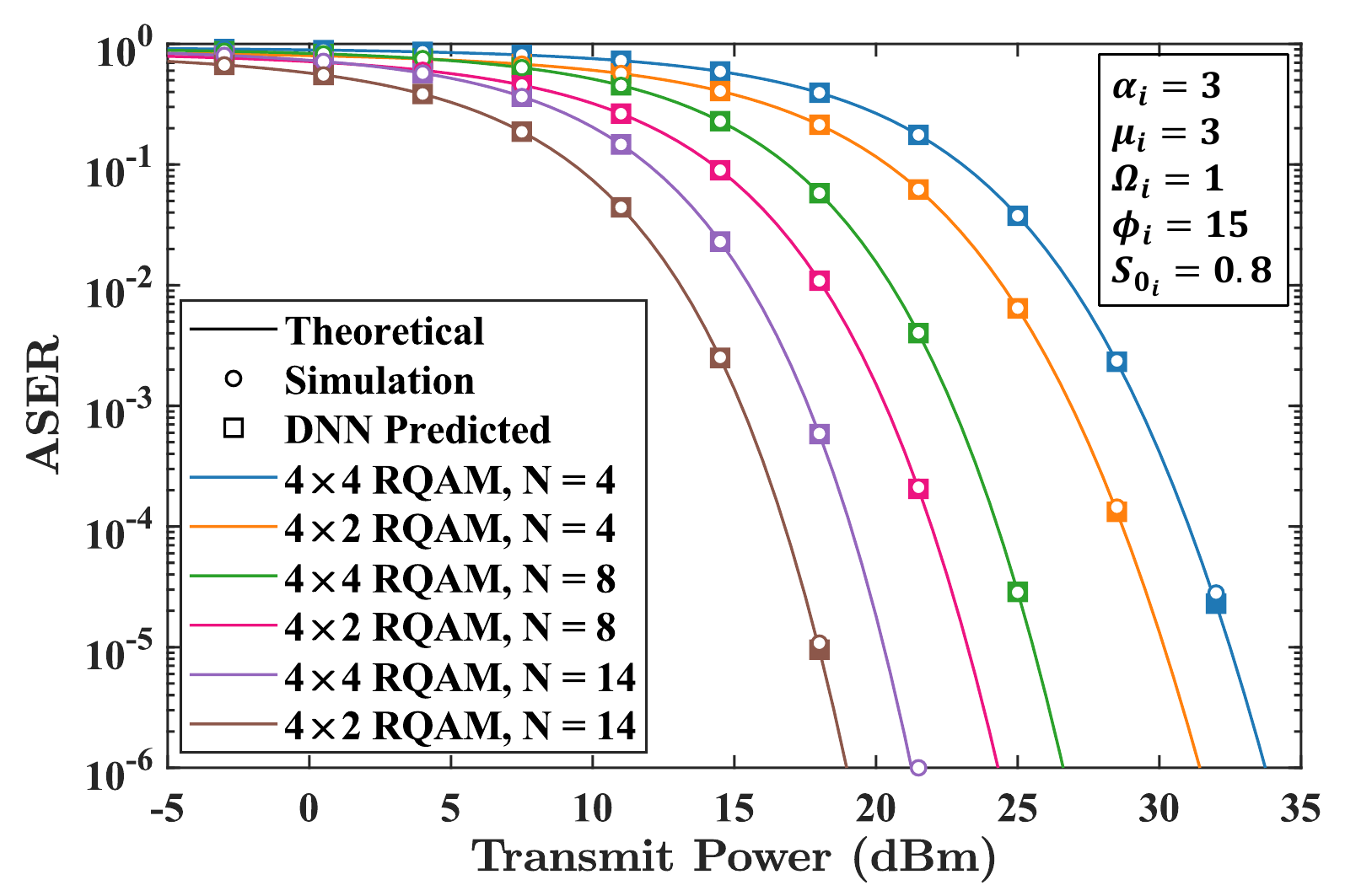}
    \caption{Impact of $M_I$, $M_Q$, and $N$ on the \ac{aser} for \ac{rqam} scheme}
    \label{fig:RQAM MI MQ N}
\end{figure}

In Fig. \ref{fig:RQAM MI MQ N}, the influence of the \ac{rqam} modulation order $M_I \times M_Q$ and the number of \ac{irs} elements $N$ on the \ac{aser} performance is illustrated. The fixed system parameters are considered as: $\alpha_i = 3$, $\mu_i = 3$, $\Omega_i = 1$, $\phi_i = 15$, and $S_{0_{i}} = 0.8$. As observed, the \ac{aser} performance deteriorates with increasing modulation order. For example, when the modulation order is increased from $4 \times 2$ to $4 \times 4$, an additional transmit power of approximately $2.3$ dB is required to achieve the same error performance across all considered values of $N = 4, 8, 14$. This degradation is attributed to the increased constellation density and reduced symbol spacing at higher modulation orders. On the other hand, a significant improvement in \ac{aser} is achieved by increasing the number of \ac{irs} elements. For instance, when $N$ is increased from $4$ to $8$ and $14$, a transmit power reduction of approximately $7.14$ dB and $12.47$ dB, respectively, is observed to maintain an \ac{aser} level of $10^{-6}$ for a fixed $4 \times 4$ \ac{rqam} scheme. This performance enhancement is due to the improved passive beamforming gain introduced by a larger number of reflecting elements. Therefore, although increasing the modulation order leads to a higher \ac{aser}, this effect can be effectively compensated by employing a sufficiently large number of \ac{irs} elements, which enhances signal quality without increasing transmit power. Finally, the predictions of \ac{aser} obtained from the \ac{dnn} model show an excellent match with the corresponding theoretical and Monte Carlo simulation results.

\begin{figure}[t!]
    \centering
    \includegraphics[width=9cm, height=6cm]{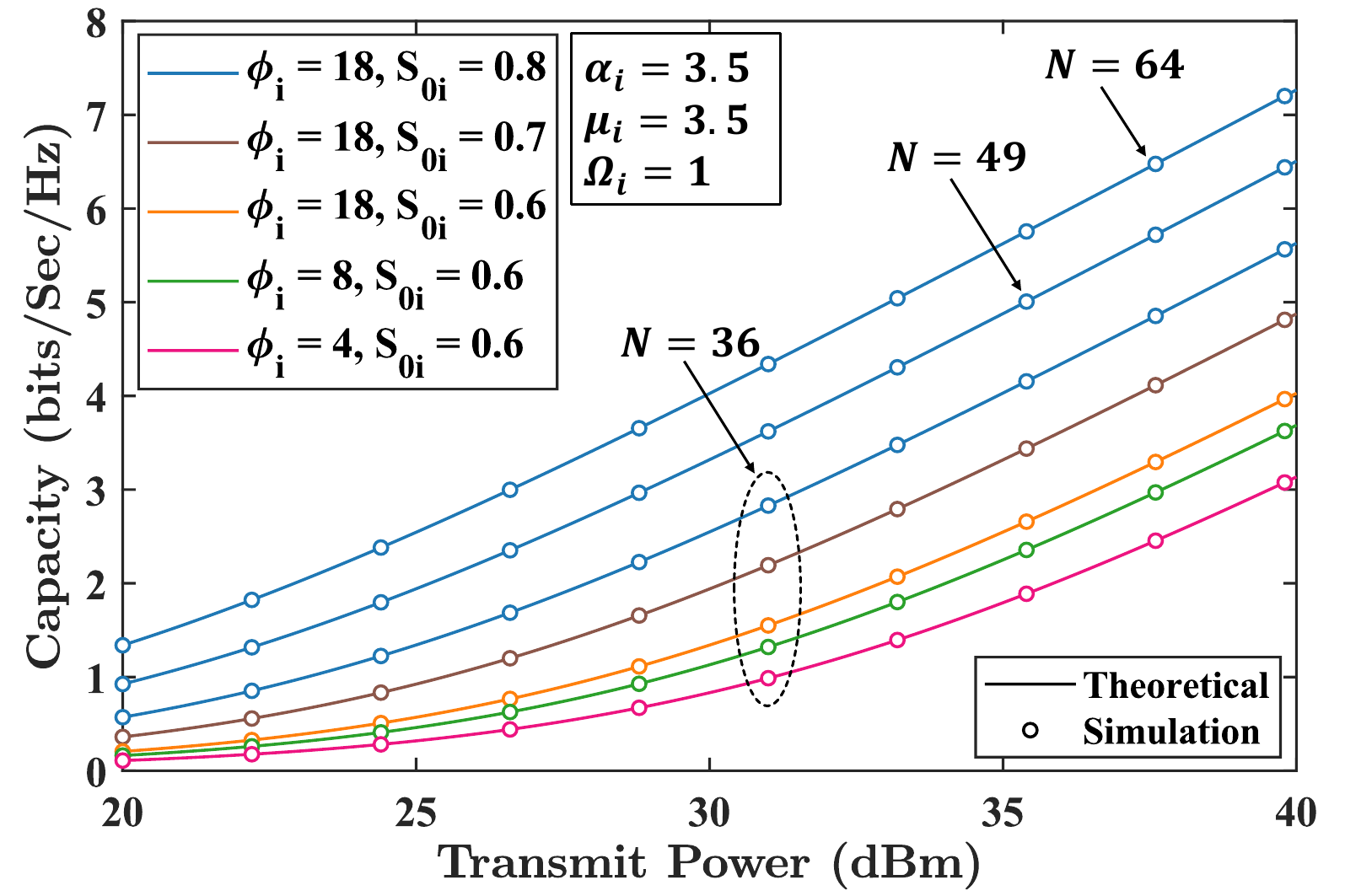}
    \caption{Impact of $\phi_i$, $S_{0i}$, and $N$ on the \ac{acc} performance}
    \label{fig:CC Phi S0}
\end{figure}

In Fig. \ref{fig:CC Phi S0}, the influence of the number of \ac{irs} elements $N$ and the pointing error parameters $\phi_i$ and $S_{0_{i}}$ on the \ac{acc} is investigated. An enhancement in the \ac{acc} is observed with increasing $N$, highlighting the role of large-scale \ac{irs} deployments. Specifically, when $N$ increases from $36$ to $49$ and subsequently to $64$, while keeping other parameters fixed as $\alpha_i = \mu_i = 3.5$, $\Omega_i = 1$, $\phi_i = 18$, $S_{0_{i}} = 0.8$, and $P_s = 38$ dBm, the \ac{acc} improves from $4.98$ to $5.85$ and $6.61$ bits/s/Hz, respectively. Additionally, the \ac{acc} exhibits sensitivity to the pointing error parameters. With $N = 36$ and $\phi_i = 18$, increasing $S_{0_{i}}$ from $0.6$ to $0.7$ and $0.8$ leads to a notable increase in \ac{acc} from $3.41$ to $4.24$ and $4.98$ bits/s/Hz, respectively, at a fixed transmit power of $38$ dBm. Moreover, the impact of $\phi_i$ is evident to achieve a target \ac{acc}. For instance, at $N = 36$ and $S_{0_{i}} = 0.6$, increasing $\phi_i$ from $4$ to $8$ and $18$ allows the system to maintain an \ac{acc} of approximately $3.13$ bits/Sec/Hz while reducing the required transmit power by approximately $1.84$ dB and $2.94$ dB, respectively. These findings demonstrate the importance of accurate beam alignment and adequate \ac{irs} deployment to achieve enhanced spectral efficiency in \ac{thz} communication systems.

\begin{figure}[t!]
    \centering
    \includegraphics[width=9cm, height=6cm]{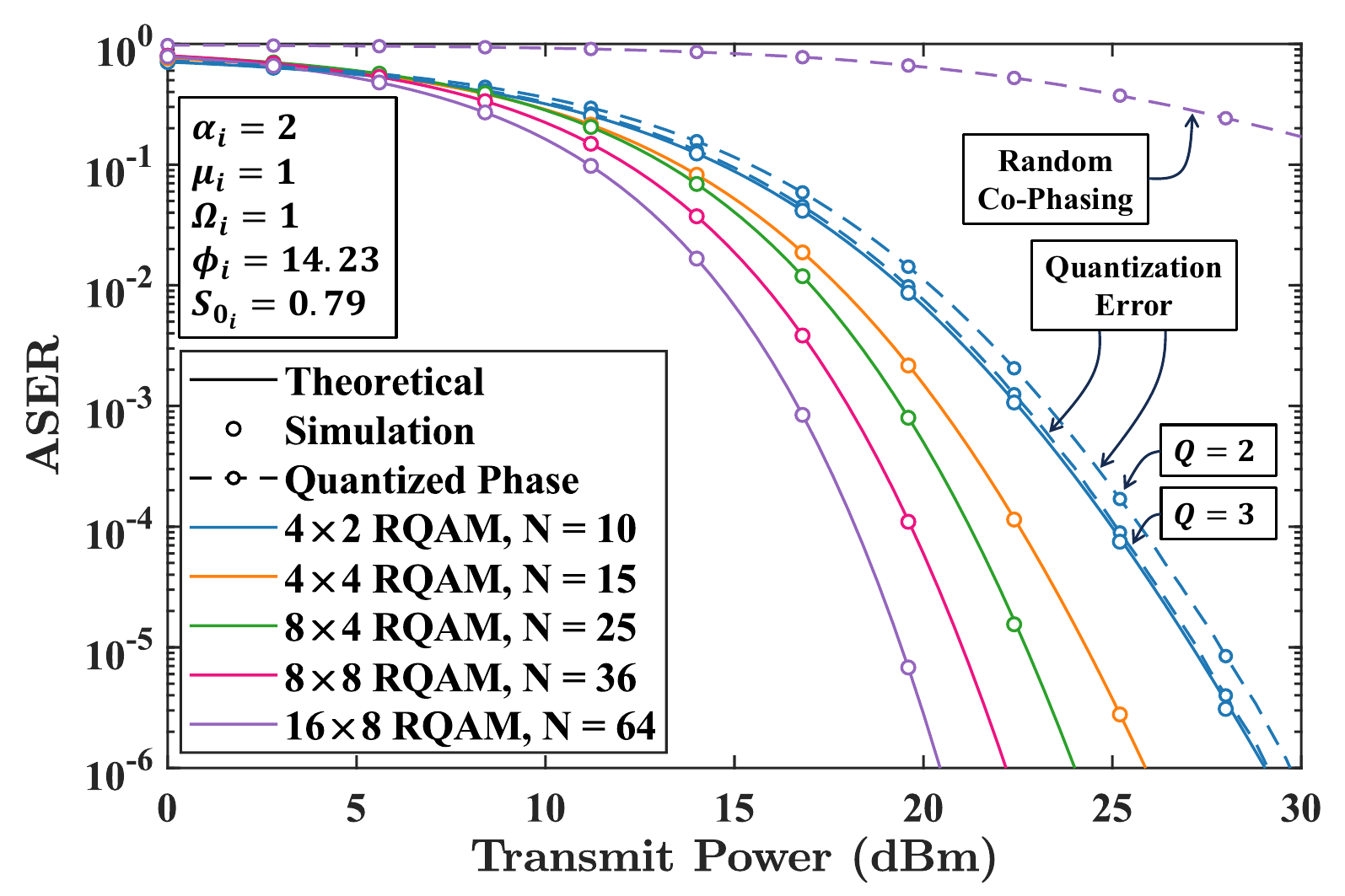}
    \caption{Impact of $N$ on the \ac{aser} performance of various higher order \ac{rqam} constellations}
    \label{fig:RQAM MI MQ Antenna}
\end{figure}
Fig.~\ref{fig:RQAM MI MQ Antenna} illustrates the influence of the number of \ac{irs} elements $N$ on the \ac{aser} performance for various \ac{rqam} configurations. The system parameters are fixed as $\alpha_i = 2$, $\mu_i = 1$, $\Omega_i = 1$, $\phi_i = 14.23$, and $S_{0_{i}} = 0.79$. The results clearly indicate that increasing $N$ effectively compensates the degradation in \ac{aser} performance typically associated with higher modulation orders. For instance, when the modulation order is increased from $4\times2$ to $4\times4$, the \ac{aser} improves from $4.73\times10^{-5}$ to $1.41\times10^{-6}$ at $P_s = 25.8$ dBm, which is solely attributed to the increase in $N$ from $10$ to $15$. Moreover, a significant performance improvement is observed when $N$ is further increased. Specifically, increasing $N$ from $10$ to $64$, approximately $8.62$ dB lower transmit power is required to achieve an \ac{aser} of $10^{-6}$, despite the modulation order being increased sixteenfold from $4\times2$ to $16\times8$. These findings highlight the essential role of large-scale \ac{irs} deployment in maintaining reliable system performance under high spectral efficiency demands, which is one of the fundamental requirements of \ac{thz} band communication. Additionally, the simulations account for both random cophasing and quantization errors. The results reveal that random cophasing leads to a considerable deterioration in the \ac{aser}. On the other hand, when the \ac{irs} employs a $Q$-bit quantizer, the impact of quantization error becomes negligible for $Q>3$, which is consistent with the trend observed in the \ac{op} performance.

\begin{figure}[t!]
    \centering
    \includegraphics[width=1\linewidth]{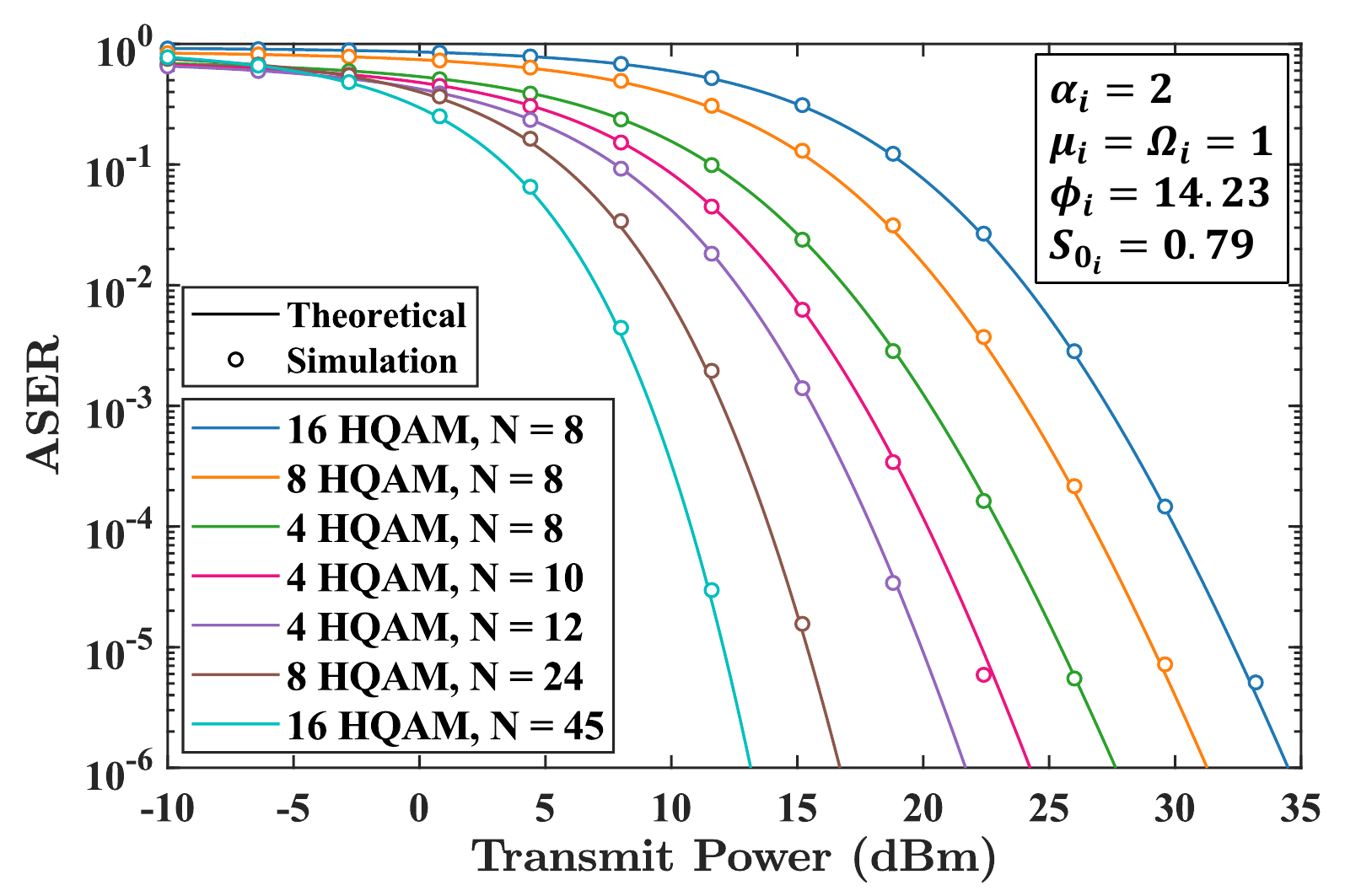}
    \caption{Impact of $M$ and $N$ on the \ac{aser} performance for the \ac{hqam} scheme}
    \label{fig:HQAM M N}
\end{figure}

Fig.~\ref{fig:HQAM M N} depicts the influence of $M$ and $N$ on the \ac{aser} performance versus transmit \ac{snr} for the \ac{hqam} scheme. The system parameters are kept constant at $\alpha_i = 2,\ \mu_i = \Omega_i = 1,\ \phi_i = 14.23,\ S_{0_{i}} = 0.79$. From the results, several key observations can be made. First, similar to the \ac{rqam} scheme, the \ac{aser} of the \ac{hqam} scheme degrades with an increase in $M$ when other parameters remain unchanged. For example, at $N = 8$, increasing $M$ from $4$ to $8$ and $16$ requires approximately $3.63$~dB and $6.84$~dB higher transmit power, respectively, to maintain an \ac{aser} of $10^{-6}$. Second, increasing the number of $N$ while keeping other parameters constant yields substantial improvements in \ac{aser} performance. Specifically, in the case of $4$-\ac{hqam}, increasing $N$ from $8$ to $10$ and $12$ reduces the required transmit power by approximately $3.39$~dB and $5.96$~dB, respectively, to achieve an \ac{aser} of $10^{-6}$. Finally, the performance degradation associated with increasing $M$ can be mitigated, or even reversed, by deploying a sufficiently large number of IRS elements. For instance, when $M$ is increased from $4$ to $8$ and $16$, and $N$ is simultaneously increased from $12$ to $24$ and $45$, the required transmit power decreases by approximately $4.99$~dB and $8.53$~dB, respectively, to maintain an \ac{aser} of $10^{-6}$.

\begin{figure}[t!]
    \centering
     \includegraphics[width=9cm, height=6cm]{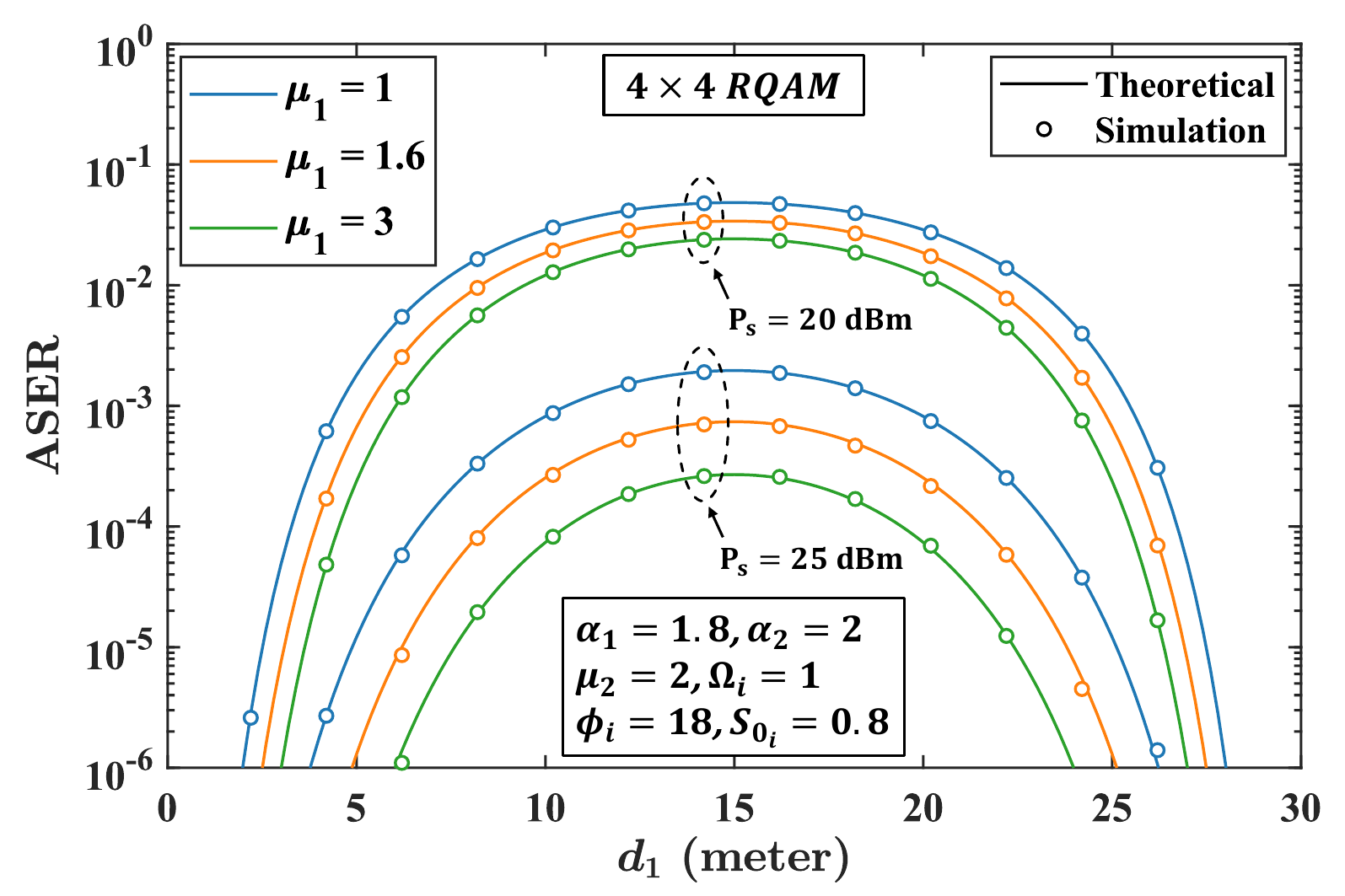}
    \caption{Impact of $d$ on the \ac{aser} performance for the \ac{rqam} scheme}
    \label{fig:RQAM Distance}
\end{figure}
Fig.~\ref{fig:RQAM Distance} illustrates the impact of the source-to-\ac{irs} distance $d_1$ on the \ac{aser} performance under an asymmetric fading scenario characterized by $\alpha_1 = 1.8$, $\alpha_2 = \mu_2 = 2$, $\Omega_i = 1$, $\phi_i = 18$, $S_{0_{i}} = 0.8$, and a fixed total distance of $d_1 + d_2 = 30$ meters. The results indicate that placing the \ac{irs} in close proximity to either \ac{s} or \ac{d} substantially enhances the \ac{aser} performance. Conversely, the highest \ac{aser} is observed when the \ac{irs} is positioned equidistant between \ac{s} and \ac{d}, despite the asymmetric fading conditions. For example, at $d_1 = d_2 = 15$ meters, increasing $\mu_1$ from 1 to 1.6 and 3 reduces the peak \ac{aser} from $1.96 \times 10^{-3}$ to $7.38 \times 10^{-4}$ and $2.69 \times 10^{-4}$, respectively, for a transmit power of $P_s = 25$ dBm. A similar trend is observed for $P_s = 20$ dBm; however, the peak becomes noticeably flatter, signifying that the improvement in \ac{aser} due to \ac{irs} placement near \ac{s} or \ac{d} is more pronounced at lower transmit power levels. Therefore, to achieve optimal symbol error performance in asymmetric \ac{thz} communication scenarios, it is advisable to deploy the \ac{irs} closer to the transceiver experiencing more severe fading.

\begin{figure}[t!]
    \centering
     \includegraphics[width=9cm, height=6cm]{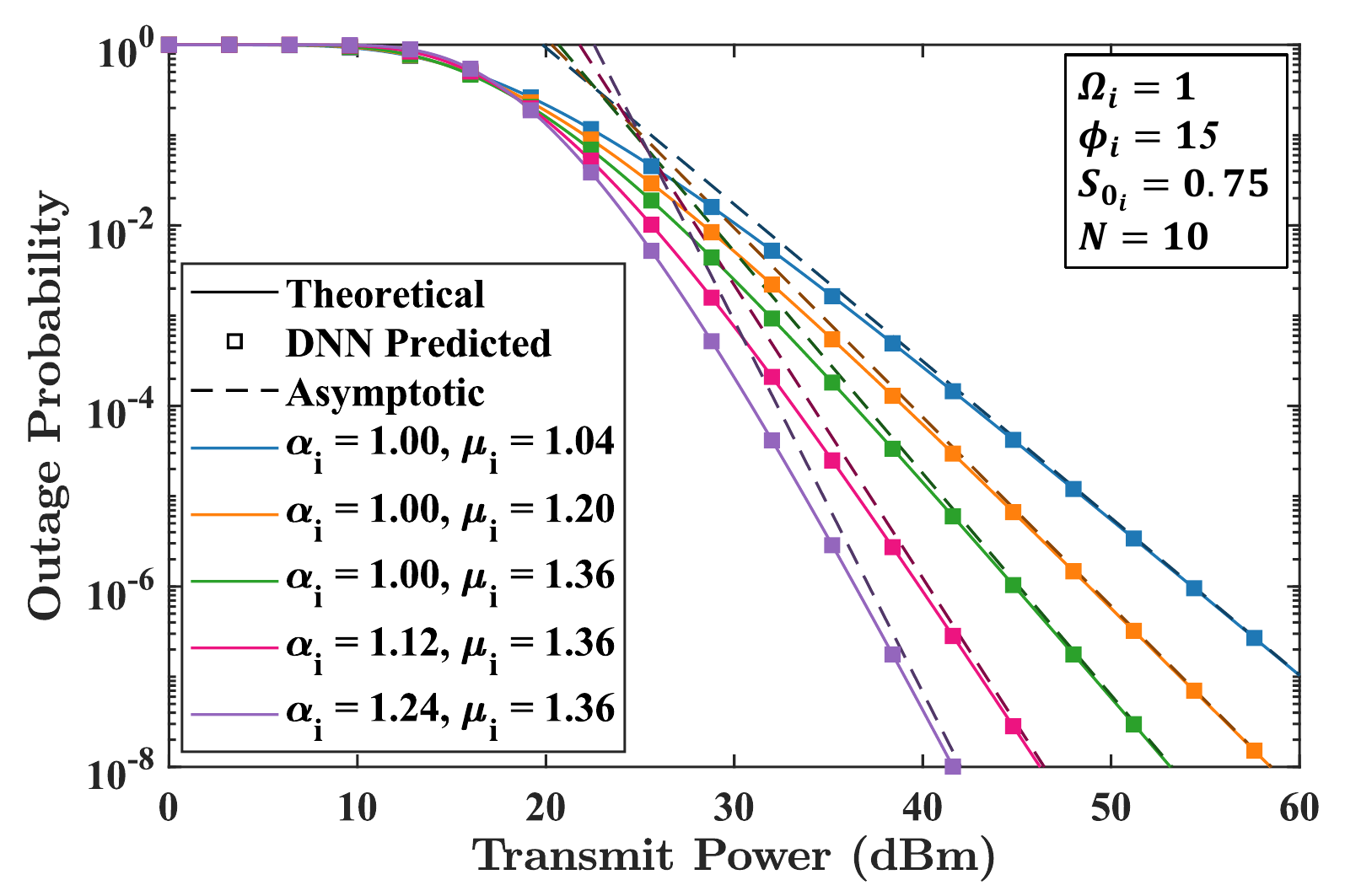}
    \caption{Diversity analysis based on asymptotic outage analysis of \ac{e2e} system}
    \label{fig:OP DO}
\end{figure}
In Fig.~\ref{fig:OP DO}, the asymptotic behavior of the \ac{op} under varying fading conditions is illustrated. Also, the \ac{dnn}-predicted \ac{op} results are presented, exhibiting substantial overlap with the theoretical results. Further, the \ac{do} analysis is performed by considering high transmit power levels (up to $P_s = 60$ dBm) for different combinations of \ac{thz} link fading parameters. Other parameters are kept fixed as: $\Omega_i = 1$, $\phi_i = 15$, $S_{0_{i}} = 0.75$, $N = 10$, and $d_1 = d_2 = 15$ m. It is evident that, with increasing transmit power, the theoretical outage curves tend to align with the asymptotic curves across all fading scenarios. This convergence confirms the validity of the asymptotic analysis in the high \ac{snr} regime. Furthermore, the slope of the \ac{op} curve in the logarithmic domain reveals the \ac{do} of the system. Specifically, for the fading parameters $\alpha_i = 1$ and $\mu_i = 1.04$, the empirical diversity order at $P_s = 60$ dBm is estimated as follows:
\begin{align*}
    &-10\left[\frac{
        \log_{10}\left(P_O(\lambda_{th})/60\,\text{dBm}\right) 
        - \log_{10}\left(P_O(\lambda_{th})/59.8\,\text{dBm}\right)
    }{
        60\,\text{dBm} - 59.8\,\text{dBm}
    }\right] \\
    &\quad= -50\log_{10}\left(\frac{1.0328 \times 10^{-7}}{1.1184 \times 10^{-7}}\right) \approx 1.729, 
\end{align*}
which is in close agreement with the theoretical \ac{do} derived from the asymptotic expression in eq.~\eqref{Simplified_Gd_Expression}:
\begin{align*}
    &\frac{10}{2}\left[\frac{\Gamma\left(3.04\right)\times\Gamma\left(1.04\right)\times16^2}{15\times 17\times\left\{\Gamma\left(2.04\right)\right\}^2}-1\right]^{-1}\approx 1.737.
\end{align*}
The close correspondence between the theoretical and empirical values corroborates the validity of the asymptotic diversity gain expression, affirming its applicability for performance prediction under high-\ac{snr} conditions.

\section{conclusion}\label{conclusion}
This work analyzed the performance of an \ac{irs}-assisted \ac{thz} band communication system using validated analytical models and extensive simulations. Closed-form analytical expressions for \ac{op}, \ac{acc}, and \ac{aser} for \ac{rqam} and \ac{hqam} schemes over the end-to-end link were derived. The results showed that increasing the fading parameters $\alpha$ and $\mu$ improved outage performance and reduced symbol error probabilities by mitigating multipath effects and lowering transmit power requirements, while the degradation in \ac{aser} caused by higher modulation orders was effectively compensated by increasing the number of \ac{irs} elements to enhance passive beamforming. It was further observed that diversity gain was determined solely by $\alpha$, $\mu$, and $\phi$. The study also highlighted that higher values of $\phi$ and $S_{0}$ improved \ac{acc}, and that optimal \ac{irs} placement preferably closer to the more severely faded transceiver enhanced \ac{aser}, confirming that large-scale \ac{irs} deployment, modulation optimization, and precise beam alignment were critical for achieving reliable and high-capacity \ac{thz} communications.
Furthermore, performance degradation due to random cophasing errors was found to be severe, whereas the impact of phase quantization error became negligible when $Q > 3$ for $Q$-bit quantizers. Finally, the proposed DNN-based framework provided a faster and efficient alternative to analytical evaluation by accurately predicting the \ac{op} and \ac{aser} across diverse system parameters, thereby offering a low-complexity solution with results closely matching theoretical analyses.
These findings strongly supported the practical implementation of \ac{thz} communication systems.

\appendix
This section provides the solutions for the integrals $\mathcal{I}_{1}(\cdot, \cdot)$, $\mathcal{I}_{2}(\cdot, \cdot)$, $\mathcal{I}_{3}(\cdot, \cdot, \cdot)$, $\mathcal{I}_{4}(\cdot, \cdot, \cdot)$, and $\mathcal{I}_{5}$ in the subsections below.

\appendices
\section{A Solution of \texorpdfstring{$\mathcal{I}_1\left(\chi_1, \chi_2\right)$}{Solution of I1(x1,x2)}}\label{Appendix_I1}
Eq. (\ref{I1_Def}) can be solved with the aid of \cite[eq. (3.381.4)]{gradshteyn2014table} as
\begin{align} \label{intLambdaE3}
    \mathcal{I}_1 \left(\chi_{1}, \chi_{2} \right) = \left( \frac{1}{\chi_2}\right)^{\chi_1 + 1} \Gamma\left(\chi_1 + 1\right).
\end{align}

\section{Solution of \texorpdfstring{$\mathcal{I}_2\left(\chi_1, \chi_2\right)$}{Solution of I2(x1,x2)}}\label{Appendix_I2}
By expressing the upper incomplete Gamma function in eq. (\ref{I2_Def}) in terms of the Meijer‑G function using \cite[eq. (06.06.26.0005.01)]{wolfram_functions}, the integral $\mathcal{I}_2\left(\chi_1, \chi_2\right)$ can be reformulated as
\begin{align}\label{I2_Step_1}
    \mathcal{I}_2\left(\chi_1, \chi_2\right) = \int_{\lambda=0}^\infty\frac{\lambda^{\chi_1}}{e^{\lambda\chi_2}}G_{1,2}^{2,0}\left[\frac{\sqrt{\lambda}}{\Lambda\sqrt{\lambda_0}}\left|\begin{array}{cc}
       1  &  \\
       \tau+1,  & 0 
    \end{array}\right.\right] d\lambda.
\end{align}
Subsequently, by employing the integral representation of the Meijer-G function \refMG, eq. (\ref{I2_Step_1}) can be rewritten as
\begin{align} \label{I2_Step_2}
    \mathcal{I}_2\left(\chi_1, \chi_2\right) = \frac{1}{2\pi i} \int_{s} \frac{\Gamma\left(\tau+1+s\right)\Gamma\left(s\right)}{\Gamma\left(1+s\right) \left(\Lambda\sqrt{\lambda_0}\right)^{-s}}\int_{\lambda=0}^\infty \frac{\lambda^{\chi_1 - \frac{s}{2}}}{e^{\lambda\chi_2}} d\lambda \ ds.
\end{align}
The inner integral in eq. (\ref{I2_Step_2}) is evaluated with the aid of \cite[eq. (3.381.4)]{gradshteyn2014table}, resulting in a simplified form of $\mathcal{I}_2\left(\chi_1, \chi_2\right)$ as
\begin{align}\label{I2_Step_3}
    \mathcal{I}_2\left(\chi_1, \chi_2\right) = \frac{1}{2\pi i} \int_{s} \frac{\Gamma\left(\tau+1+s\right)}{s \left(\Lambda\sqrt{\lambda_0}\right)^{-s}} \frac{\Gamma\left(1+\chi_1-\frac{s}{2}\right)}{\chi_2^{1+\chi_1-\frac{s}{2}}} ds.
\end{align}
Finally, by utilizing \refFoxH, eq. (\ref{I2_Step_3}) can be expressed in closed form in terms of the Fox’s H-function as
\begin{align}\label{I2_Step_4}
    \mathcal{I}_2\left(\chi_1, \chi_2\right) = \frac{1}{\chi_2^{\chi_1+1} }H_{2,2}^{2,1}\left[\frac{1}{\Lambda\sqrt{\chi_2\lambda_0}}\left|\begin{array}{cc}
         \nu_0\left(\chi_1\right), \nu_1\\
         \nu_2, \nu_3
    \end{array}\right.\right], 
\end{align}
where, $\nu_0\left(\chi\right)=\left(-\chi, \frac{1}{2}\right)$, $\nu_1=\left(1,1\right)$, $\nu_2 = \left(\tau+1, 1\right)$, and $\nu_3 = \left(0,1\right)$.

\section{A Solution of \texorpdfstring{$\mathcal{I}_3 \left(\chi_{1}, \chi_{2} \right)$}{Solution of I3(x1,x2)}}\label{Appendix_I3}
The confluent hypergeometric function ${}_1F_1(\cdot;\cdot;\cdot)$ appearing in eq. (\ref{I3_Def}) can be expressed in its series form with the aid of \cite[eq. (9.210.1)]{gradshteyn2014table}. Accordingly, the integral $\mathcal{I}_3 \left(\chi_{1}, \chi_{2}, \chi_{3}\right)$ can be written as
\begin{align} \label{intE1F1_2}
    \mathcal{I}_3 \left(\chi_{1}, \chi_{2}, \chi_{3}\right) = \sum_{\eta=0}^{\infty} \frac{\Gamma\left(\frac{3}{2}\right)\Gamma\left(\eta+1\right)}{\Gamma\left(\frac{3}{2}+\eta\right)}\frac{\chi_3^\eta}{\eta!}\int_{\lambda = 0}^{\infty} \frac{\lambda^{\eta+\chi_1}}{e^{\chi_{2} \lambda}} d\lambda.
\end{align}
With the aid of \cite[eq. (3.381.4)]{gradshteyn2014table}, eq. (\ref{intE1F1_2}) can be evaluated as
\begin{align} \label{intE1F1_3_1}    
    \mathcal{I}_3 \left(\chi_{1}, \chi_{2}, \chi_{3}\right) =  \sum_{\eta=0}^{\infty} &\frac{\Gamma\left(\eta+1\right)\Gamma\left(\chi_1+\eta+1\right)\Gamma\left(\frac{3}{2}\right)}{\Gamma\left(\chi_1+1\right)\Gamma\left(\frac{3}{2}+\eta\right)}  \nonumber \\ \times
    & \frac{\Gamma\left(\chi_1+1\right)}{\chi_2^{\chi_1+1}} \frac{\chi_3^\eta}{\chi_2^\eta\eta!}.
\end{align}
Finally, eq. (\ref{intE1F1_3_1}) can be expressed in terms of the Gaussian hypergeometric function using \cite[eq. (7.2.3)]{prudnikov} as
\begin{align} \label{intE1F1_4}\textbf{}
    \mathcal{I}_3 \left(\chi_{1}, \chi_{2}, \chi_{3}\right) = \frac{\Gamma\left(\chi_1+1\right)}{\chi_2^{\chi_1+1}} \, {}_{2}F_{1}\left(1,\chi_1+1;\frac{3}{2};\frac{\chi_3}{\chi_2}\right).
\end{align}

\section{Solution of \texorpdfstring{$\mathcal{I}_4\left(\chi_1, \chi_2, \chi_3\right)$}{Solution of I4(x1, x2, x3)}}\label{Appendix_I4}
With the aid of \cite[eqs. (06.06.26.0005.01) and (07.20.26.0015.01)]{wolfram_functions}, the upper incomplete Gamma function and the confluent hypergeometric function in eq. (\ref{I4_Def}) are expressed in terms of Meijer‑G functions, allowing $\mathcal{I}_4\left(\chi_1, \chi_2, \chi_3\right)$ to be written as
\begin{align}\label{I4_Intermediate_1}
    \mathcal{I}_4\left(\chi_1, \chi_2, \chi_3\right)=\frac{1}{2}\int_{\lambda=0}^\infty &\frac{\lambda^{\chi_1}}{e^{\left(\chi_2-\chi_3\right)\lambda}}G_{1,2}^{1,1}\left[\chi_3\lambda\left|\begin{array}{cc}
        0.5  \\
        0, -0.5
    \end{array}\right.\right] \nonumber \\
    &\times G_{1,2}^{2,0}\left[\frac{\sqrt{\lambda}}{\Lambda\sqrt{\lambda_0}}\left|\begin{array}{cc}
         1\\
         \tau+1, 0 
    \end{array}\right.\right] d\lambda.
\end{align}
Further, by employing the integral representation of the Meijer-G function \refMG, $I_4\left(\chi_1, \chi_2, \chi_3\right)$ can be reformulated as
\begin{align} \label{I4_Intermediate_2}
    I_4\left(\chi_1,\right.&\left.\chi_2, \ \chi_3\right) = \frac{1}{2}\left(\frac{1}{2\pi i}\right)^2\int_r\int_s\frac{\Gamma\left(r\right)\Gamma\left(0.5-r\right)}{\Gamma\left(1.5-r\right)\chi_3^r} \times \nonumber \\
    &\frac{\Gamma\left(\tau+1+s\right)\Gamma\left(s\right)}{\Gamma\left(1+s\right)\left(\Lambda\sqrt{\lambda_0}\right)^{-s}} \int_{\lambda = 0 }^\infty \frac{\lambda^{\chi_1-r-\frac{s}{2}}}{e^{\left(\chi_2-\chi_3\right)\lambda}} d\lambda \ dr \ ds.
\end{align}
Furthermore, the inner integral in eq. (\ref{I4_Intermediate_2}) can be evaluated with the aid of \cite[eq. (3.381.4)]{gradshteyn2014table}, yielding
\begin{align} \label{I4_Intermediate_3}
    I_4\left(\chi_1,\right.&\left.\chi_2, \ \chi_3\right) = \frac{1}{2}\left(\frac{1}{2\pi i}\right)^2\int_r\int_s\frac{\Gamma\left(r\right)\Gamma\left(0.5-r\right)}{\Gamma\left(1.5-r\right)\chi_3^r} \times \nonumber \\
    &\frac{\Gamma\left(\tau+1+s\right)\Gamma\left(s\right)}{\Gamma\left(1+s\right)\left(\Lambda\sqrt{\lambda_0}\right)^{-s}} \frac{\Gamma\left(1+\chi_1-r-\frac{s}{2}\right)}{\left(\chi_2-\chi_3\right)^{1+\chi_1-r-\frac{s}{2}}} \ dr \ ds.
\end{align}
Finally, using the integral definition of the multivariate Fox–H function \refMVFoxH, $\mathcal{I}_4\left(\chi_1, \chi_2, \chi_3\right)$ can be expressed in closed form as
\begin{align} \label{I4_Final}
    I_4\left(\chi_1, \chi_2, \chi_3\right) = \mathcal{C}_{1}H_{1,0:1,2;1,2}^{0,1:1,1;2,0}\left[\begin{array}{cc}
        \nu_4  \\
        \nu_5
             \end{array}\left|\begin{array}{c}
                \nu_6\left(\chi_1\right):\nu_7;\nu_1\hfill  \\
                -:\nu_3, \nu_8;\nu_2,\nu_3 
             \end{array}\right.\right],
\end{align}
where $\mathcal{C}_1 = \frac{1}{2\left(\chi_2 - \chi_3\right)^{1 + \chi_1}}$, $\nu_4 = \frac{\chi_3}{\chi_2 - \chi_3}$, $\nu_5 = \frac{1}{\Lambda \sqrt{\left(\chi_2 - \chi_3\right)\lambda_0}}$, $\nu_6(\chi) = \left(-\chi, 1, \frac{1}{2}\right)$, $\nu_7 = \left(0.5, 1\right)$, and $\nu_8 = \left(-0.5, 1\right)$.

\section{A Solution of \texorpdfstring{$\mathcal{I}_{5}$}{Solution of I5}}\label{Appendix_I5}
By using the integral form representations of the Meijer‑G function from \refMG $\,$ and the upper incomplete Gamma function from \cite[eq. (8.350.2)]{gradshteyn2014table} in eq. (\ref{I5_Def}), the expression for $\mathcal{I}_{5}$ can be written as
\begin{align}\label{I5_intermediate_3}
    \mathcal{I}_{5} = \frac{1}{2\pi i} \int_{s}\Gamma\left(s\right)\Gamma\left(1-s\right)\int_{\lambda=0}^\infty\lambda^{-s} \int_{t=\frac{\sqrt{\lambda}}{\Lambda\sqrt{\lambda_0}}}^\infty \frac{t^{\tau}}{e^t} dt \ d\lambda \ ds.
\end{align}
Further, by interchanging the order of integration with respect to the variables $\lambda$ and $t$, the integral $\mathcal{I}_{5}$ can be expressed as
\begin{align}\label{I5_intermediate_4}
    \mathcal{I}_{5} = \frac{1}{2 \pi i}&\int_s \Gamma\left(s\right)\Gamma\left(1-s\right)\int_{t=0}^\infty \frac{t^\tau}{e^t} \int_{\lambda=0}^{\Lambda^2 \lambda_{0} t^2} \frac{d\lambda}{\lambda^{s}} \ dt \ ds.
\end{align}
By evaluating the inner integral with respect to $\lambda$, eq. (\ref{I5_intermediate_4}) can be rewritten as
\begin{align}\label{I5_intermediate_5}
    \mathcal{I}_{5} = \frac{1}{2 \pi i}\int_{s} \frac{\Gamma\left(s\right)\Gamma\left(1-s\right)}{\left(1-s\right)\left(\Lambda^2\lambda_0\right)^{s-1}} \int_{t=0}^{\infty} e^{-t} t^{\tau+2-2s} \ dt \ ds.
\end{align}
Furthermore, by using \refGamma, the inner integral in eq. (\ref{I5_intermediate_5}) can be expressed in terms of the Gamma function, yielding
\begin{align}\label{I5_intermediate_6}
    \mathcal{I}_{5}=\frac{1}{2 \pi i}\int_{s}\frac{\Gamma\left(s\right)\left\{\Gamma\left(1-s\right)\right\}^2\Gamma\left(3+\tau-2s\right)}{\Gamma\left(2-s\right)\left(\Lambda^2\lambda_0\right)^{s-1}} ds.
\end{align}
Now, considering $r=s-1$, the integral in eq. \eqref{I5_intermediate_6} can be expressed with respect to $r$ as
\begin{align}\label{I6_intermediate_6}
    \mathcal{I}_{5}=\frac{1}{2 \pi i}\int_{r}\frac{\Gamma\left(r+1\right)\left\{\Gamma\left(-r\right)\right\}^2\Gamma\left(1+\tau-2r\right)}{\Gamma\left(1-r\right)\left(\Lambda^2\lambda_0\right)^{r}} dr.
\end{align}

Finally, by utilizing the integral definition of the Fox's H-function \refFoxH, the expression for $\mathcal{I}_5$ can be expressed as
\begin{align}\label{I5_Final}
    \mathcal{I}_{5} =  H_{3,2}^{1,3}\left[\Lambda^2\lambda_0\left|\begin{array}{cc}
         \nu_1, \nu_1, \nu_9  \\
         \nu_1, \nu_3
    \end{array}\right.\right],
\end{align}
where, $\nu_9 = \left(-\tau, 2\right)$.

\bibliographystyle{IEEEtranDOI}  
\bibliography{References.bib}

\end{document}